\documentclass[12pt]{article}
\DeclareUnicodeCharacter{0302}{\^}
\usepackage[utf8]{inputenc}
\usepackage{graphicx}
\usepackage{natbib}
\usepackage{color}
\usepackage{amsmath}
\usepackage{amsfonts}
\usepackage{hyperref}
\usepackage[algo2e]{algorithm2e} 
\usepackage{algorithm}
\usepackage{verbatim}
\usepackage{caption}
\usepackage{setspace}
\usepackage{subcaption}
\usepackage{multirow}
\usepackage[left=1in,right=1in,top=1in,bottom=1in]{geometry}

\newcommand{\bx}{\boldsymbol{x}}

\newcommand{\bbeta}{\boldsymbol{\beta}}

\title{Pseudo-R2D2 prior for high-dimensional ordinal regression}
\author{Eric Yanchenko \\ Akita International University}

\begin{document}

\maketitle
\thispagestyle{empty}


\begin{abstract}
\noindent
Ordinal regression with a high-dimensional covariate space has many important application areas including gene expression studies. The lack of an intrinsic numeric value associated with ordinal responses, however, makes methods based on continuous data, like linear regression, inappropriate. In this work, we extend the R2D2 prior framework to the high-dimensional ordinal setting. Since the $R^2$ definition used in the R2D2 prior relies on means and variances, it cannot be used for ordinal regression as these two quantities are not suitable for such data. Instead, by simulating data and using McFadden's coefficient-of-determination ($R^2_M$), we show that a generalized inverse Gaussian prior distribution on the global variance parameter approximately induces a beta prior distribution on $R^2_M$. The proposed prior can be implemented in \texttt{Stan} and an $\texttt{R}$ package is also developed. Our method demonstrates excellent inference properties on simulated data, as well as yielding accurate predictions when applied to a liver tissue gene expression dataset. 

\end{abstract}


\section{Introduction}\label{sec:intro}
Ordinal data arises in many applications, making it an important problem for regression. A hallmark of ordinal responses is that they have a natural ordering to them, but do not inherently correspond to a numeric value. Because of this, typical regression models that rely on continuous variables are not suitable, and may lead to incorrect conclusions \citep[e.g.][]{stevens1946theory}. Moreover, in the modern big data era, high-dimensional data sets also occur quite often, leading to the ``small $n$, large $p$'' setting, where there are a small number of responses but a large number of associate covariates. Such problems have garnered significant research interest for continuous, binary and count response data, especially in the Bayesian community, but these methods are not directly applicable to ordinal regression settings. 

To fix ideas, consider the following three examples where ordinal data appear. Perhaps the most common example is survey response data, particularly related to psychological experiments \citep{burkner2019ordinal}. In this setting, participants may be shown a statement, and asked to give their opinion on a scale from ``strongly disagree,'' ``moderately disagree,'' ``moderately agree,'' or ``strongly agree.'' While these response data have a clear ordering, there is no inherent numeric value that corresponds to, e.g., ``moderately disagree.'' Moreover, the difference between ``strongly disagree'' and ``moderately disagree'' is not necessarily the same as the difference between ``moderately agree'' and ``strongly agree.''

While survey data may be the most common situation where ordinal data arise, there are also numerous examples from industrial experiments \citep{chipman1996bayesian}. For example, in judging the appearance of steel after some experiment, it may be too difficult and/or costly to develop a continuous scale of the appearance. A steel expert may instead grade the appearance is ``poor,'' ``fair,'' or ``good.'' As in the survey data, there is no obvious numeric value that can be assigned to the responses.  

Lastly, and of particular interest in this present work, is the occurrence of high-dimensional ordinal response data in medical applications. For example, cancerous tumors are typically rated on the ordinal scale of Phase I to Phase IV. Additionally, in hepatocellular carcinoma (HCC) studies, liver tissues may be ordered from normal to hepatitis C virus (HCV) infected but no HCC to HCV infected with HCC \citep{archer2010high, archer2010identifying, zhang2021bayesian}. In these examples, there may only be tens or hundreds of patients enrolled in the study, but for a single patient, we may have access to thousands of covariates corresponding to their gene expression. Thus, it is important to have ordinal regression models that can perform well both in terms of inference and prediction on such datasets.

There has been some previous work on high-dimensional ordinal regression models. The first contributions come from the frequentist literature where $L_1$ \citep{archer2012} or Elastic-Net \citep{wurm2021regularized} penalties are used to shrink the coefficient estimates towards zero. Recently, \cite{zhang2021bayesian} proposed a Bayesian solution. The authors adopt ideas from both the Bayesian LASSO \citep{tibshirani1996regression, park2008bayesian, hans2009bayesian}, as well as spike-and-slab prior frameworks \citep{mitchell1988bayesian}  to develop a variable selection procedure. In particular, the authors endow the regression coefficients with a double exponential prior, while also including a binary parameter to encode the inclusion/exclusion of the coefficient in the model.

In any Bayesian hierarchical model, setting the prior for scale parameters is a major challenge. As the models have multiple levels, it is not always clear what the effect of the prior choice will be on parameters in different levels, nor is there always a good intuition for choosing sensible hyper-parameters. Recently, the R2D2 prior \citep{zhang2022bayesian} has proven to be a shrinkage prior framework that gives a principled and interpretable method for assigning priors to scale parameters. Focusing on the linear regression setting, the authors first define a Bayesian $R^2$ as a ratio of the variance of the mean function to the variance of the response. By endowing this $R^2$ with a beta distribution, this induces a beta prime distribution on the global variance parameter. The resulting prior was shown to have excellent theoretical and empirical properties, and due to its success, it has been extended to generalized linear models \citep{yanchenko2024r2d2, aguilar2024generalized}, spatial regression model \citep{yanchenko2024spatial}, survival models \citep{feng2024mediation}, and more. 

In this work, we are interested in extending the R2D2 paradigm to ordinal regression. Unfortunately, this approach cannot be used directly due to the nature of ordinal data. As mentioned before, ordinal data does not have an inherent numerical value associate with it, making approaches like multiple linear regression inappropriate. For the same reason, it would also be unsuitable to use an $R^2$ definition that relies on means and variances, as in e.g., \cite{zhang2021bayesian, yanchenko2024r2d2}. 

To address these challenges, we propose an extension of the R2D2 paradigm to ordinal data which yields desirable shrinkage properties on the regression coefficients. After detailing why the original R2D2 prior is not appropriate in this setting, we leverage similar modeling ideas, but base the approach on McFadden's coefficient-of-determination ($R^2_M$) definition \citep{mcfadden1974}. By simulating data and assuming a generalized inverse Gaussian (GIG) distribution \citep{seshadri2012inverse} on the global variance parameter, we compute the hyperparameters which approximately induce a beta prior distribution on $R^2_M$. Furthermore, by leveraging an auxiliary variable formulation for the GIG distribution, we fit this model in \texttt{Stan} \citep{carpenter2017}, and develop a user-friendly \texttt{R} package. We apply the proposed method to simulated data and show that it outperforms its competitors in terms of inference properties and run-time. Finally, when applied to a hepatocellular carcinoma (HCC) liver tissue study, the proposed method's predictive ability is demonstrated. This paper represents an exciting development in the shrinkage prior literature, as we adopt the R2D2 paradigm but apply it to a new measure of model fit, opening the door to future innovations.

The road map of the paper is as follows. In Section \ref{sec:method}, we present the main methodological contribution of the paper, the $\mathsf{pR2D2ord}$ prior. This is then applied to both simulated and real world data in Sections \ref{sec:sim} and \ref{sec:real}, respectively, before closing with concluding thoughts in Section \ref{sec:conc}.

\section{Methodology}\label{sec:method}
In this section, we define the data-generating model before presenting the proposed prior framework.

\subsection{Data-generating model}
Let $Y_i$ be the $i$th (ordinal) response where $Y_i\in\{1,2,\dots,K\}$ and $i\in\{1,2,\dots,n\}$ such that $K$ is the number of response categories, and $n$ is the total number of observations.\footnote{Even though it may seem that we are assigning a numeric value to the ordinal response, in what follows this is used to simply order the responses, and the values themselves are not important.} Additionally, let $\bx_i=(x_{i1},\dots,x_{ip})^\top$ be the covariates and $\bbeta=(\beta_1,\dots,\beta_p)^\top$ the regression coefficients. We assume that $\bx_1,\dots,\bx_n\stackrel{\text{iid.}}{\sim}({\bf 0}_p,{\bf V})$ where ${\bf 0}_p$ is a vector of zeros of length $p$ and all diagonal entries of ${\bf V}$ are equal to 1. Then the linear predictor, $\eta_i$, is
\begin{equation}
    \eta_i = \bx^\top_i\bbeta.
\end{equation}

In this work, we assume that the ordinal responses are related to the linear predictors via the {\it cumulative} or {\it cut-point} model \citep[e.g.,][]{burkner2019ordinal}. In particular, each response has an associated latent variable, $\tilde Y_i$, following a normal distribution centered on the linear predictor, i.e., for all $i$,
\begin{equation}\label{eq:latent}
    \tilde Y_i\mid \eta_i\stackrel{\text{ind.}}{\sim}\mathsf{Normal}(\eta_i,1).
\end{equation}
Additionally, with $K$ response categories, we assume that there are $K-1$ cut-points, $\boldsymbol{\tau}=\tau_1,\dots,\tau_{K-1}$ which relate the latent variable to the observed response. Specifically, 
\begin{equation}\label{eq:resp}
    Y_i=
    \begin{cases}
        1,&\tilde Y_i<\tau_1\\
        k, & \tau_{k-1}\leq \tilde Y_i < \tau_k,\ k=2,\dots,K-1\\
        K, & \tau_{K-1}\leq \tilde Y_i.
    \end{cases}
\end{equation}
Thus the probability mass function is described as, 
\begin{equation}\label{eq:pmf}
    P(Y=k\mid \eta)=\Phi(\tau_k-\eta)-\Phi(\tau_{k-1}-\eta),\ k=1,\dots,K
\end{equation}
where $\Phi(\cdot)$ is the distribution function for the standard normal, $\tau_0=-\infty$ and $\tau_K=\infty$.

Since we propose a Bayesian solution, we must also specify a model for $\bbeta$. Following \cite{zhang2022bayesian, yanchenko2024r2d2}, we let $\beta_j|\phi_j,W\stackrel{\text{ind.}}{\sim}\mathsf{Normal}(0,\phi_j W)$ for $j\in\{1,\dots,p\}$, where $W>0$ corresponds to the global variance of the coefficients, and $\phi_j\geq0$ satisfy $\sum_{j=1}^p\phi_j=1$ and apportion the variance to the individual coefficients. It is common to model $\boldsymbol{\phi}$ with the Dirichlet distribution \citep{zhang2022bayesian, yanchenko2024r2d2}, although other extensions are possible \citep{aguilar2024generalized}. Regardless of the choice of prior for $\boldsymbol{\phi}$, it is easy to show that
\begin{equation}\label{eq:eta}
    \eta\mid W\sim\mathsf{Normal}(0,W)
\end{equation}
where we have dropped the index $i$ as the observations are exchangeable. We can also integrate $\eta$ out of the distribution of $\tilde Y$ and find
$$
    \tilde Y\mid W \sim \mathsf{Normal}(0,1+W)
$$
such that
\begin{equation}\label{eq:pmf}
    P(Y=k\mid W)= \Phi_W(\tau_{k})-\Phi_W(\tau_{k-1}),\ k=1,\dots,K
\end{equation}
where $\Phi_W(\cdot)$ is the distribution function for a normal random variable with mean 0 and variance $1+W$. Then our complete prior framework is:

\begin{align*}
    \beta_j|\phi_j, W&\sim \mathsf{Normal}(0,\phi_jW)\\
    \boldsymbol{\phi}&\sim\mathsf{Dirichlet}(\xi_0,\dots,\xi_0)\\
    W, \tau_1,\dots,\tau_k&\sim \pi(\cdot)
\end{align*}
It remains to specify prior distributions for $\boldsymbol{\tau}$ and $W$.

\subsection{Prior for cut-points}
We first derive a prior distribution for $\boldsymbol{\tau}$. One common approach for modeling the cut-points is to assume a sorted normal distribution, i.e., $\tau_1,\dots,\tau_{K-1}\sim\mathsf{Normal}(0,\sigma^2_\tau)$ where $\tau_1<\tau_2<\cdots<\tau_{K-1}$ \citep[e.g.,][]{archer2022ordinalbayes}. A major drawback of this choice, however, is that it is not straightforward to incorporate prior domain knowledge into the hyperparameter $\sigma^2_\tau$. Indeed, as $\boldsymbol{\tau}$ controls the probability of the response values, if a practitioner expects the responses to be, e.g., small, then it would be desirable to encode that into our hyperparameter choices.

Motivated by this desire, we propose the following prior distribution on the cut-points \citep{betancourt2019}. Let $\pi_k^W$ be the {\it a priori} probability that a response equals $k$, i.e., $P(Y=k|W)=\pi_k^W$ for $k=1,\dots,K$ and $\sum_{k=1}^K\pi_k^W=1$, where we explicitly include $W$ in the notation to emphasize its dependence. A natural choice for $\boldsymbol{\pi}^W$ is again the Dirichlet distribution, i.e., $\boldsymbol{\pi}^W=(\pi_1^W,\dots,\pi_K^W)\sim\mathsf{Dirichlet}(\alpha_1,\dots,\alpha_K)$. Based on our model in \eqref{eq:pmf}, the relationship between $\boldsymbol{\pi}^W$ and $\boldsymbol{\tau}$ is
\begin{equation}\label{eq:pi_tau}
    \pi_k^W(\boldsymbol{\tau})
    =\begin{cases}
        \Phi_W(\tau_1)&k=1\\
        \Phi_W(\tau_k)-\Phi_W(\tau_{k-1})&k=2,\dots,K-1\\
        1-\Phi_W(\tau_{K-1})&k=K
    \end{cases}
\end{equation}
where $\Phi_W(\cdot)$ is the distribution function of a normal random variable with mean 0 and variance $1+W$. Given the Dirichlet distribution of $\boldsymbol{\pi}^W$ and \eqref{eq:pi_tau}, it is a straightforward change of variables to obtain the induced distribution of $\boldsymbol{\tau}$:
\begin{equation}\label{eq:tau_pdf}
    f(\boldsymbol{\tau})
    =d(\boldsymbol{\pi}^W(\boldsymbol{\tau});\boldsymbol{\alpha})\times|{\bf J}|,\ \tau_1<\tau_2<\dots<\tau_{K-1}.
\end{equation}
where $d(\cdot;\boldsymbol{\alpha})$ is the probability density function of the Dirichlet distribution with concentration parameters $\boldsymbol{\alpha}=(\alpha_1,\dots,\alpha_K)^T$, and ${\bf J}=(J_{jk})$ is the $K-1\times K-1$ matrix defined by:
$$
    J_{jk}
    =\begin{cases}
        \phi_W(\tau_j)&k=j\\
        -\phi_W(\tau_j)&k=j-1\\
        0&\text{otherwise}
    \end{cases}
$$
where $\phi_W(\cdot)$ is the probability distribution function of a normal random variable with mean 0 and variance $1+W$.
We refer the interested reader to the Supplemental Materials for further details on the derivation.

Given the relationship between $\boldsymbol{\pi}^W$ and $\boldsymbol{\tau}$, it is straightforward to  incorporate prior domain knowledge into our prior for the cut-points. Indeed, as $\boldsymbol{\alpha}$ is the concentration parameter for the Dirichlet distribution on $\boldsymbol{\pi}^W$, we can set these values to reflect our prior belief. For example, if $K=3$ and we expect that many response values will be small, we could set $\boldsymbol{\alpha}=(5,1,1)$ to favor smaller response values. On the other hand, the prior belief that each response is equally likely would induce a ``vague'' prior where each concentration parameter $\alpha_k$ has the same value. Indeed, the relative values of $\alpha_k$ encode our relative belief about the probability of the response being equal to $k$.

\begin{figure}
    \centering
    \includegraphics[width=0.99\linewidth]{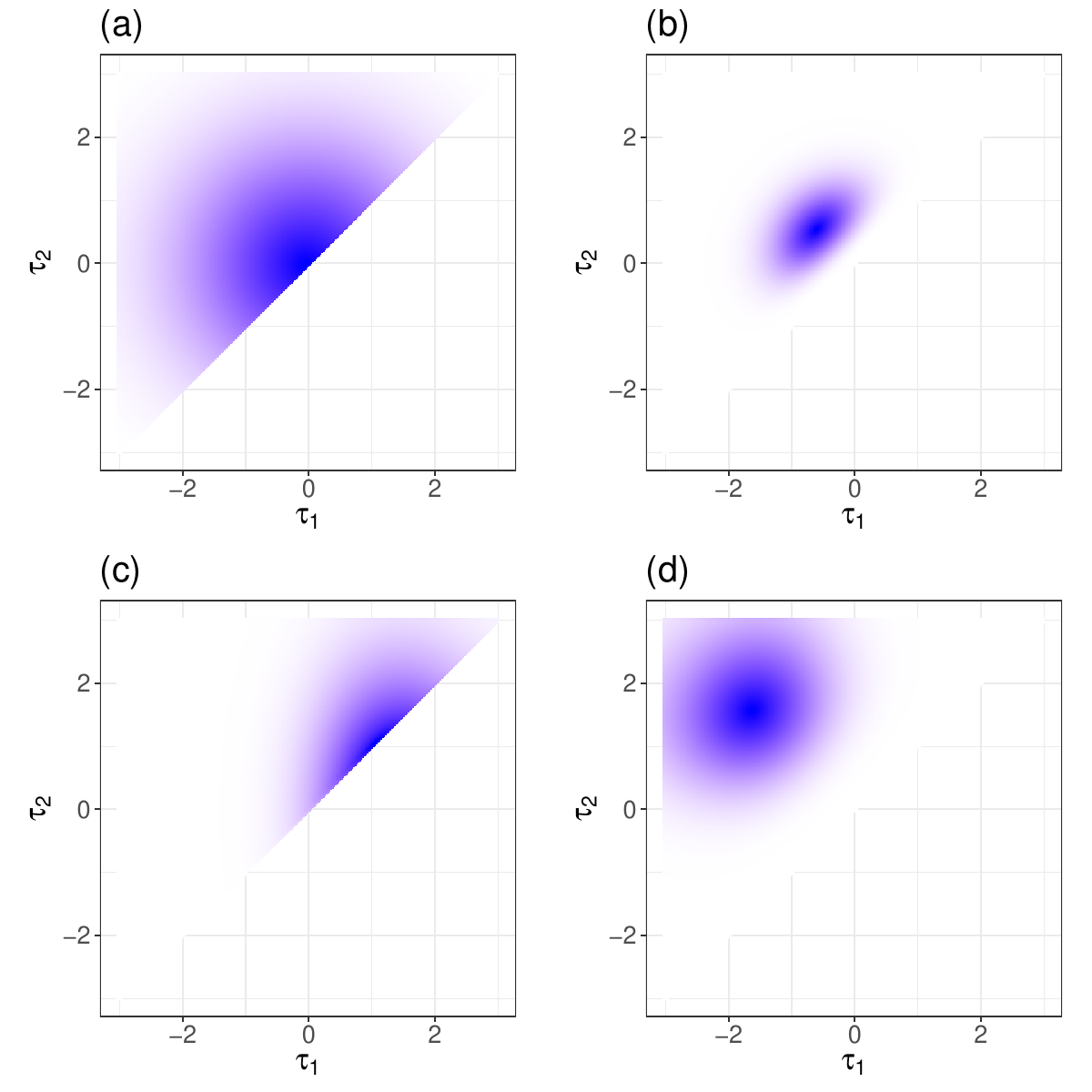}
    \caption{Prior distribution of $\boldsymbol{\tau}$ for $W=1$ and $\boldsymbol{\alpha}=(1,1,1)$ (a), $(5,5,5)$ (b), $(5,1,1)$ (c), $(1,5,1)$ (d).}
    \label{fig:tau1}
\end{figure}

Since \eqref{eq:tau_pdf} is non-standard, we plot its probability distribution function for different values of $\boldsymbol{\alpha}$ and $W$. In the following figures, we set $K=3$ such that there are two cut-points, $\boldsymbol{\tau}=(\tau_1,\tau_2)$ to enable plotting in two dimensions. In Figure \ref{fig:tau1}, we plot the probability distribution function from \eqref{eq:tau_pdf} for fixed $W=1$ and, for panels (a) - (d), the concentration parameters $\boldsymbol{\alpha}$ are $(1,1,1)$, $(5,5,5)$, $(5,1,1)$ and $(1,5,1)$, respectively. Note that the color corresponds to the probability density such that regions with darker blue are more likely.

\begin{figure}
    \centering
    \includegraphics[width=0.99\linewidth]{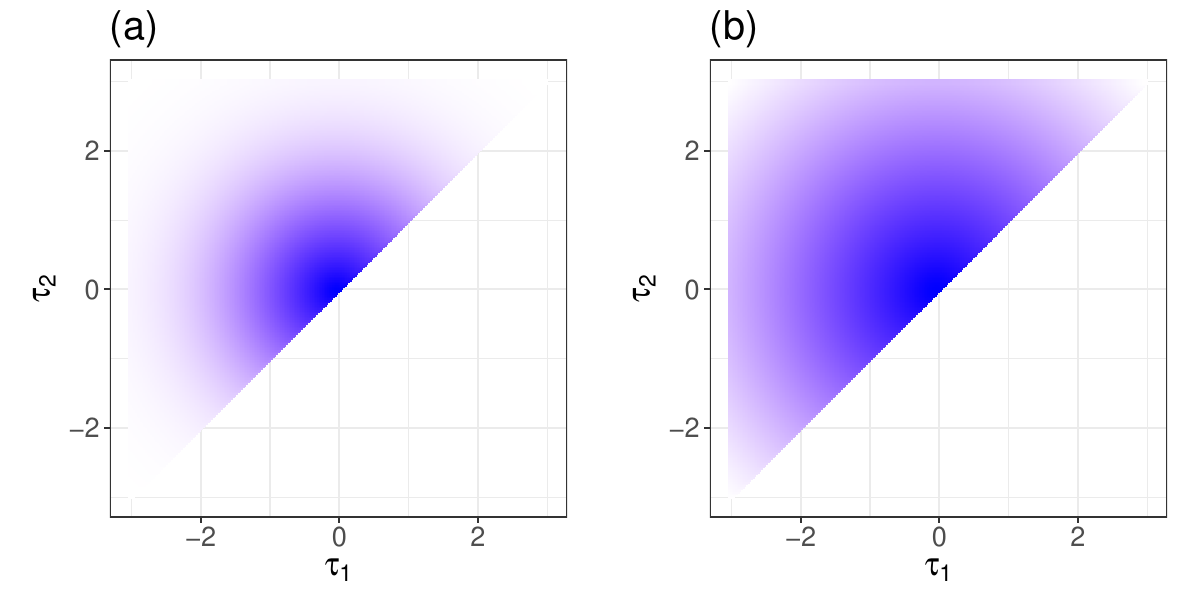}
    \caption{Prior distribution of $\boldsymbol{\tau}$ for $\boldsymbol{\alpha}=(1,1,1)$ and $W=0.5$ (a) and $W=5$ (b).}
    \label{fig:tau2}
\end{figure}

In panels (a) and (b), we have the same concentration parameter for all $k$ which encodes a symmetric distribution of $\boldsymbol{\tau}$. In panel (b), however, the larger values of $\alpha_k$ correspond to smaller variance in the Dirichlet distribution for $\boldsymbol{\pi}^W$ which in turn yields smaller variance (smaller blue region) for $\boldsymbol{\tau}$. The choice of $\boldsymbol{\alpha}=(5,1,1)$ in (c) encodes the prior belief that many response values will be 1. Indeed, if $\tau_1$ and $\tau_2$ are both likely to be large, as suggested in the figure, then this favors response values of 1 as $\Phi_W(\tau_1)$ will be large while $\Phi_W(\tau_2)-\Phi_W(\tau_1)$ and $1-\Phi_W(\tau_2)$ will both be small. Similarly, in (d) where $\boldsymbol{\alpha}=(1,5,1)$, the plot shows a high probability of small $\tau_1$ and large $\tau_2$, making $\Phi_W(\tau_2)-\Phi_W(\tau_1)$ large, encouraging response values of two. Lastly, in Figure \ref{fig:tau2}, we fix $\boldsymbol{\alpha}=(1,1,1)$ and vary $W=0.5$ (a) and $W=5$ (b). We see that larger values of $W$ induce larger variation in the distribution for $\boldsymbol{\tau}$.

\subsection{Bayesian $R^2$}
Given the distribution of $\boldsymbol{\tau}$, we turn our attention to the primary goal of this work, finding a distribution of $W$ which yields good shrinkage properties for $\bbeta$. Motivated by the R2D2 paradigm \citep[e.g.][]{zhang2022bayesian}, we first must define a Bayesian coefficient-of-determination ($R^2$). Given a response $Y$ and linear predictor $\eta$, the Bayesian $R^2$ for a generalized linear model is
\begin{equation}\label{eq:r2}
    R^2
    =\frac{\mathsf{Var}\{\mu(\eta)\}}{\mathsf{Var}\{\mu(\eta)\} + \mathsf{E}\{\sigma^2(\eta)\}} 
\end{equation}
where $\mathsf{E}(Y|\eta)=\mu(\eta)$ and $\mathsf{Var}(Y|\eta)=\sigma^2(\eta)$ \citep{yanchenko2024r2d2}. Then we seek a distribution of $W$ which induces a prior distribution on $R^2$ with desired properties, i.e., $R^2\sim\mathsf{Beta}(a,b)$.

In principle, it is straightforward to extend this definition to the ordinal regression model. Given \eqref{eq:pmf}, the expectation of the response variable is
\begin{equation}\label{eq:ord_mean}
    \mathsf{E}(Y|\boldsymbol{\tau}, W)
    =\sum_{k=1}^K k\{\Phi_W(\tau_k)-\Phi_W(\tau_{k-1})\},
\end{equation}
and similar logic yields an expression for the variance. We can substitute these results into \eqref{eq:r2} and leverage the ideas of \cite{yanchenko2024r2d2} to obtain the generalized beta prime distribution for $W$ which induces a desirable distribution on $R^2$.

The astute reader, however, will notice that the $R^2$ definition in \eqref{eq:r2} is quite unsatisfying from a statistically philosophical angle as this definition requires the mean and variance of the response variable. While these quantities can be computed for ordinal data, it is well-documented that such measures are, in fact, inappropriate \citep[e.g.,][]{stevens1946theory, stevens1955averaging, liddell2018analyzing}. As \cite{stevens1946theory} argues, referring to ordinal data, ``In the strictest propriety the ordinary statistics involving means and standard deviations ought not to be used with these scales, for these statistics imply a knowledge of something more than the relative rank-order of data... means and standard deviations computed on an ordinal scale are in error to the extent that the successive intervals on the scale are unequal in size.'' In other words, because the response data refer to ranked categories and not strictly to an equidistance scale, any measure which assumes such a property in the data (like the mean and variance) is inappropriate.

To illustrate this point, consider response data from a survey, where participants are asked to rate their subjective experience after taking a drug, from $\{$\texttt{poor}, \texttt{fair}, \texttt{good}, \texttt{excellent}$\}$. While we may encode that \texttt{poor} corresponds to 1, \texttt{fair} to 2, etc., by computing the mean and variance of these responses, we are implicitly assuming that the {\it distance} between each response is the same; e.g., the difference between the experience of \texttt{poor} and \texttt{fair} is the same as the difference between \texttt{good} and \texttt{excellent}. While some argue that we can ignore this philosophical quandary and simply proceed with caution in using means and variances for ordinal data, we will show that the proposed approached is not only more satisfying philosophically, but also leads to improved empirical performance.

\subsection{McFadden's $R^2$}
To circumvent the aforementioned philosophical difficulties, while still leveraging the R2D2 paradigm, we need an $R^2$ definition appropriate for ordinal data. One such measure is  {\it McFadden's} $R^2$ \citep{mcfadden1974}. Originally derived as a post-hoc measure of model fit, McFadden's $R^2$ ($R^2_M$) is a pseudo-$R^2$, defined as
\begin{equation}\label{eq:mc}
    R^2_M=1-\frac{\log L_M}{\log L_0},
\end{equation}
where $L_M,L_0$ correspond to the likelihood of the full and null models, respectively. From \eqref{eq:pmf}, the joint likelihood for the full model is
\begin{equation}\label{eq:like}
    L_M
    =\prod_{i=1}^n \prod_{k=1}^K \{\Phi_W(\tau_k)-\Phi_W(\tau_{k-1})\}^{\mathbb I(Y_i=k)},
\end{equation}
where $\mathbb I(\cdot)$ is the indicator function. Given the responses ${\bf Y}=(Y_1,\dots,Y_n)^\top$, cut-points $\boldsymbol{\tau}$, and global variance $W$, we can calculate the log-likelihood in \eqref{eq:like}.

To complete the specification of $R^2_M$, however, we also need a notion of the null model. Recall that the linear predictor is $\eta_i=\bx_i^\top\bbeta$ where $\beta_j|\phi_j, W\stackrel{\text{ind.}}{\sim}\mathsf{Normal}(0,\phi_jW)$. If $W=0$, then $\eta=0$, encoding the ``intercept-only'' model\footnote{We note that our model framework does not include an intercept because this parameter is unnecessary given the latent variable and cut-point model.}. In other words, the null model corresponds to all regression coefficients being 0, which accords with the notion of a null model \citep[e.g.,][]{simpson2017penalising}. If $W=0$, then the null likelihood is
\begin{equation}\label{eq:like0}
    L_0
    =\prod_{i=1}^n \prod_{k=1}^K \{\Phi(\tau_k)-\Phi(\tau_{k-1})\}^{\mathbb I(Y_i=k)}.
\end{equation}
Given \eqref{eq:like} and \eqref{eq:like0}, we can compute $R^2_M$.

This coefficient-of-determination does not depend on the mean or variance of the response, but only its probability of occurring. In this way, it serves as a valid measure of model fit for ordinal data. There are many other measures of model fit that are appropriate \citep[e.g.,][]{nagelkerke1991note, cohen2013applied, allison2014measures}, but we leave their investigations for future work.

\subsection{Setting prior on $R^2_M$}
The goal of the R2D2 paradigm is to find a distribution for $W$ which induces a desirable distribution on $R^2_M$. Here, we seek $R^2_M\sim\mathsf{Beta}(a,b)$ for a user-defined $a$ and $b$. In the previous sub-section, we showed that $R^2_M$ is a suitable measure of model fit for ordinal data, but there remains a major hurdle in implementing this definition for prior specification. Based on \eqref{eq:mc} - \eqref{eq:like0}, the response values ${\bf Y}$ are needed in order to calculate the likelihoods, and therefore $R^2_M$ as well. Since the prior should be set {\it before} observing the data, this difficulty must be addressed.

\subsubsection{Data simulation}
To side-step this challenge, we propose simulating a prior distribution of $R^2_M$. In particular, after choosing $\boldsymbol{\alpha}$, consider the following procedure:
\begin{enumerate}
    \item Sample $W\sim \pi(\cdot)$
    \item Sample $\boldsymbol{\tau}\sim f(\boldsymbol{\tau};\boldsymbol{\alpha})$ from \eqref{eq:tau_pdf}
    \item Sample $\tilde Y_1,\dots,\tilde Y_n\mid W\stackrel{\text{iid.}}{\sim}\mathsf{Normal}(0,1+W)$
    \item Compute ${\bf Y}$ using \eqref{eq:resp}
    \item Compute $R^2_M$
    \item Repeat steps 1 - 5 a large number of times to obtain a prior distribution of $R^2_M$
\end{enumerate}

By following these steps, we can construct an empirical prior distribution for $R^2_M$, without using the observed response variables, and only requiring specification of $n$, $K$ and $\boldsymbol{\alpha}$. Thus, all that remains is to set the prior distribution for global variance parameter, $W$.

\subsubsection{Generalized Inverse Gaussian distribution}
We propose using the {\it Generalized Inverse Gaussian (GIG)} \citep{seshadri2012inverse} distribution as the prior distribution for $W$. The GIG distribution is a flexible three-parameter family of distributions which often arises in Bayesian context as it conjugate for the distribution of the global variance, and has both the gamma and inverse-gamma distributions as special cases \citep[e.g.,][]{bhattacharya2015dirichlet, zhang2022bayesian}. If $X\sim\mathsf{GIG}(\lambda,\rho,\chi)$, then for $x>0$ and $\rho,\chi>0$,
$$
    f(x)
    =\frac{(\rho/\chi)^{\lambda/2}}{2K_\lambda(\sqrt{\rho\chi)}}x^{\lambda-1}e^{(-\rho x+\chi/x)/2}
$$
where $K_\lambda(\cdot)$ is the modified Bessel function of the second kind. Notice that the GIG distribution reduces to the gamma and inverse gamma distribution if $\chi=0$ and $\rho=0$, respectively. 

\subsubsection{Loss function}
For a user-selected $a,b$, we propose finding the hyper-parameters $\lambda^*,\rho^*,\chi^*$ such that if $W\sim\mathsf{GIG}(\lambda^*,\rho^*,\chi^*)$, and we generate $R^2_M$ using the procedure above, then the resulting distribution will be approximately $\mathsf{Beta}(a,b)$ distributed. We measure the distance between two distributions using the {\it 2-Wasserstein metric}, i.e., we seek to minimize
\begin{equation}\label{eq:obj}
    D^2_{(a,b)}(\lambda,\rho,\chi)=\int_0^1 \{ \tilde F^{-1}_{R^2_M}(\tau; \lambda,\rho,\chi) - F^{-1}(\tau;a,b)\}^2 d\tau
\end{equation}
where $\tilde F^{-1}_{R^2_M}(\cdot; \lambda,\rho,\chi)$ is the empirical quantile function of $R^2_M$, and $F^{-1}(\tau;a,b)$ is the quantile function of a $\mathsf{Beta}(a,b)$ distributed random variable. In practice, we approximate this integral with a sum and find
\begin{equation}\label{eq:dist}
    \lambda^*,\rho^*,\chi^2
    =\arg\min_{\lambda,\rho,\chi} D^2_{(a,b)}(\lambda,\rho,\chi).
\end{equation}
Other divergence measures were also considered (e.g., Kullback-Liebler, Pearson $\chi^2$) but we found that the 2-Wasserstein metric performed best in practice. Lastly, to ensure the optimization routine does not get stuck in a local minima, we suggest running it multiple times and keeping the result which minimizes the objective function in \eqref{eq:obj}.

\subsubsection{Full procedure}
To summarize, for a given $\boldsymbol{\alpha},\ n,\ K,\ a,$ and $b$, we generate a distribution of $R^2_M$ using the procedure above, and compute its distance from a $\mathsf{Beta}(a,b)$ distribution using \eqref{eq:dist}. Then using an optimization procedure, we minimize this distance to obtain the optimal hyperparameters $\lambda^*,\rho^*,\chi^2$ such that if $W\sim\mathsf{GIG}(\lambda^*,\rho^*,\chi^*)$, then $R^2_M\stackrel{\text{approx.}}{\sim}\mathsf{Beta}(a,b)$. 

In the Supplemental Materials, we plot histograms which show the relationship between the empirical distribution of $R^2_M$ using the optimal hyper-parameters, and the desired beta distribution. In general, we show a good correspondence between the observed and theoretical distributions, with the closest match occurring for distributions with large prior mass of $R^2_M$ near 0. Indeed, we find it difficult to obtain values of $R^2_M$ near (but not equal to) 1. For example, the fit for the $\mathsf{Beta}(1,5)$ distribution is better than for the $\mathsf{Beta}(5,1)$. This is not a significant issue for at least two reasons. First, as discussed in \cite{yanchenko2024r2d2}, $R^2_M=0$ (which corresponds to $W=0$), can be considered as the ``base-model'' for this framework. In this way, large prior mass near $R^2_M=0$ ($W=0$) encourages shrinkage in the regression coefficient estimates, {\it a la} the penalized complexity prior \citep{simpson2017penalising}. Indeed, this was the base model used in our definition of $R^2_M$. Thus, we suggest choosing values of $a$ and $b$ corresponding to such distributions. Second, the R2D2 paradigm gives a principled framework for guiding prior distribution choices and/or hyperparameter selection. Thus, even if the prior distribution for $R^2_M$ is not exactly $\mathsf{Beta}(a,b)$, the resulting hyperparameters can be used, and the model still yields good posterior distribution properties. Finally, in the Supplemental Material, we present a table of the optimal hyperparameter values of GIG prior for various values of $a,b,K$.

\subsubsection{Auxiliary variable formulation}
Before writing the final prior framework, we note one final challenge to overcome. Ideally, this model could be implemented in \texttt{Stan} \citep{carpenter2017} to ensure that it is easy for practitioners to use.  \texttt{Stan} cannot implement the GIG distribution, however, because it does not support the modified Bessel function of the second kind for non-integer $\lambda$. To implement this model, we rely on an auxiliary variable formulation for the GIG distribution from \cite{pena2025}. Specifically, if $\lambda>-1/2$,\footnote{The case for $\lambda\leq \frac12$ is also available, but in practice we found that the optimization procedure almost always yields values of $\lambda > -1/2$, so for the sake of space we only present this case.} then iteratively sampling from
\begin{align}\label{eq:w_aux}
    W\mid\xi&\sim\mathsf{InvGauss}({(\chi+2\xi)/\rho}^{1/2}, \chi+2\xi)\\\notag
    \xi\mid W&\sim \mathsf{Gamma}(\lambda+1/2, 1/w)
\end{align}
is equivalent to sampling from $W\sim\mathsf{GIG}(\lambda,\rho,\chi)$, where $\mathsf{InvGauss}(\mu,\sigma)$ is the Inverse Gaussian distribution with mean $\mu>0$, and shape parameter $\sigma>0$. Unlike the GIG distribution, the Inverse Gaussian distribution can be implemented in \texttt{Stan}. 

\subsection{Complete framework}

We now present the entire prior specification.
\begin{align*}
    \beta_j|\phi_j, W&\sim \mathsf{Normal}(0,\phi_jW)\\
    \boldsymbol{\phi}&\sim\mathsf{Dirichlet}(\xi_0,\dots,\xi_0)\\
    \boldsymbol{\tau}&\sim f(\boldsymbol{\tau};\boldsymbol{\alpha})\\
    W\mid\xi&\sim\mathsf{InvGauss}(\{(\chi^*+2\xi)/\rho^*\}^{1/2}, \chi^*+2\xi)\\
    \xi\mid W&\sim \mathsf{Gamma}(\lambda^*+1/2, 1/W)
\end{align*}

\noindent
Since we followed the R2D2 paradigm, but used a pseudo-$R^2$ measure of model fit, we coin this the $\mathsf{pR2D2ord}$ prior.

\subsection{Discussion}
We close this section with a brief discussion of the $\mathsf{pR2D2ord}$ prior. First, we extended the R2D2 framework to a different $R^2$ definition (McFadden's) which was suitable for ordinal data. This extension was not trivial, as it required a data simulation scheme to approximate the prior distribution. After selecting the GIG distribution for $W$, we proposed an optimization scheme based on the 2-Wasserstein metric to obtain the hyperparameters which most closely yield the desired prior distribution of $R^2_M$. Finally, an auxiliary variable formulation allows this model to be implemented in \texttt{Stan}.

One of the major advantages of the proposed prior framework is that it allows two different ways to incorporate domain knowledge, as well as automatic prior selection in the absence of such information. First, if there is prior knowledge on the response values, then this information can be incorporate into the values of $\boldsymbol{\alpha}$. For example, larger values of $\alpha_k$ encode the prior belief that the event $Y=k$ is more likely to occur. Since $R^2_M$ is an interpretable measure of model fit, it is also straightforward to incorporate prior domain knowledge via the  hyperparameters $a,b$. If the practitioner believes that the model will fit the data well, then a larger prior $R^2_M$ could be preferred, i.e., $a=5,b=1$. On the other hand, if the data is expected to be quite noisy, perhaps a smaller prior $R^2_M$ is preferable, i.e., $a=1,b=5$. In the absence of such domain knowledge, the $\mathsf{pR2D2ord}$ prior easily allows for automatic hyperparameter selection as well. For example, $a=b=1$ constitutes an agnostic belief on the model fit, while significant prior mass near 0 should be the default choice for high-dimensional problems to enforce sparsity in the model. Either way, the proposed method yields both an intuitive framework to incorporate domain knowledge, as well as an automatic procedure.

Finally, we develop the \texttt{R2D2ordinal} package, available on \texttt{CRAN}, to implement the proposed method in \texttt{R}. In addition to a user-friendly implementation of this model in \texttt{Stan}, we also provide functions to compute the probability distribution function of the cut-points, $f(\boldsymbol{\tau};\boldsymbol{\alpha})$, compute the log-likelihood \eqref{eq:like}, generate the distribution of $R^2_M$, and find the optimal hyperparameters for the GIG distribution.

\section{Simulation study}\label{sec:sim}
In this section, we demonstrate the performance of the proposed $\mathsf{pR2D2ord}$ prior on synthetic data. In particular, we are interested in its ability to perform inference on $\bbeta$.

\subsection{Settings}
We generate data using the following scheme. First, we generate $n$ vectors of covariates $\bx\sim\mathsf{MVN}({\bf 0}_p,{\bf V})$ where ${\bf 0}_p$ is a vector of zeros of length $p$, and ${\bf V}$ is an AR(1) covariance matrix with correlation $\rho=0.8$. Next, we generate (sparse) $\bbeta$ coefficients under two different schemes (discussed below) and sample the latent variable $\tilde Y|\eta\sim\mathsf{Normal}(\eta,1)$ where $\eta_i=\bx_i^\top\bbeta$. After selecting the number of response categories $K$, we set the cut-points $\boldsymbol{\tau}=(\tau_1,\dots,\tau_{K-1})$, and given $\tilde{\bf Y}$, we find the response values ${\bf Y}$.

There are several parameters in the data-generating process that we can vary. First, we consider four different combinations of $n$ and $p$, i.e., 
$$
    (n,p)\in\{(50,100),\ (50,250),\ (100,500),\ (100,1000)\},
$$
and two different number of responses categories, $K=3,5$.
We also vary how the coefficients $\bbeta$ are generated. For all values of $p$, we set $6$ coefficients to be non-null ($\neq0$) and the remaining $p-6$ to be 0. For the non-null coefficients, we either set these to be $1$ or $-1$ (denoted, ``Fixed'' in all results), or generate them from a $t_3$ distribution ($t_3$). Finally, to vary the distribution of the response values, we use two different methods to set $\boldsymbol{\tau}$. To enforce the same number of response values in each category, we set $\tau_k$ as the $k/K$th quantile of $\tilde{\bf Y}$ (Even). For an imbalanced setting with more small response values, we set $\tau_1=0$ and $\tau_k=k/K+1$ for $k>2$ (Low). This results in 32 total simulation settings. In the main text, we present the results for $(n,p)\in\{(50,250),\ (100,500)\}$ (16 settings), and the remaining 16 settings are left to the Supplemental Materials. Each setting is repeated for $R=100$ Monte Carlo (MC) replications and the average values are reported.

\subsection{Evaluation Metrics}
We are primarily interested in the inference properties of the proposed method. First, we compute the mean-squared error between the true value of the coefficients $\bbeta$ and the posterior median, i.e.,
$$
    \text{mse} = \frac1{pR}\sum_{r=1}^R\sum_{j=1}^p (\beta_j-\hat\beta^r_j)^2
$$
where $\hat\beta_j^r$ is the posterior median for coefficient $j$ for the $r$th MC replicate. To assess the variable selection properties, we also compute the area under the receiver-operator curve (AUC) using the absolute value of the posterior median $|\hat{\bbeta}|$ as the ``predictions'' and the null/non-null status of $\beta_j$ as the response. We evaluate the frequentest coverage rate of the 95\% credible intervals, as well as compute its width, i.e.,
$$
    \text{cov}
    =\frac1{pR}\sum_{r=1}^R\sum_{j=1}^p \mathbb I\{\hat\beta_j^r(0.025)\leq \beta_j\leq \hat\beta_j^r(0.975)\}
$$
and 
$$
    \text{width}
    = \frac1{pR}\sum_{r=1}^R\sum_{j=1}^p \{\hat\beta_j^r(0.975) - \hat\beta_j^r(0.025)\},
$$
where $\hat\beta_j^r(q)$ is the $q$th quantile of the posterior distribution if the $r$th replicate. Lastly, we report the computational time.


\subsection{Competing Methods}
We compare the proposed $\mathsf{pR2D2ord}$ prior with several competitors. While not originally designed for ordinal regression, we can implement the original R2D2 prior \citep{zhang2022bayesian} and R2D2glm prior \citep{yanchenko2024r2d2}. In both cases, the prior specification is identical to the one discussed here, save the distribution of $W$. For the R2D2 prior, we take $W$ to have a beta prime distribution, i.e.,
$$
    W\sim\mathsf{BetaPrime}(a,b),
$$
while for the R2D2glm, it takes a generalized beta prime distribution
$$
    W\sim\mathsf{GBP}(a^*,b^*,c^*,d^*).
$$
Please see \cite{yanchenko2024r2d2} for more details on computing the hyper-parameters $(a^*,b^*,c^*,d^*)$. The R2D2 prior can be thought of as inducing a $\mathsf{Beta}(a,b)$ prior distribution on $R^2$ using the latent mean, $\eta$. For R2D2glm, on the other hand, this prior induces an approximate $\mathsf{Beta}(a,b)$ distribution on the marginal $R^2$ in \eqref{eq:r2}, but note that this definition uses the mean and variance of the ordinal responses. Both of these models are implemented in \texttt{Stan}.

We consider two other competing methods. \cite{zhang2021bayesian} proposed a shrinkage prior for ordinal regression, based on the idea of the spike-and-slab LASSO prior \citep[e.g.,][]{george1993, rovckova2018spike}. Specifically, this model, which we call ssLASSO, assumes
\begin{align*}
    \beta_j&= \gamma_j Z_j\\
    Z_j&\sim \mathsf{DE}(\lambda)\\
    \lambda&\sim \mathsf{Gamma}(a_0,b_0)\\
    \gamma_j&\sim \mathsf{Bernoulli}(\pi_0)\\
    \tau_k&\sim\mathsf{Normal}(0,\tau_0^2)\text{ where }\tau_1<\tau_2<\cdots<\tau_{K-1}.
\end{align*}
where $\mathsf{DE}(\lambda)$ is the double exponential (Laplace) distribution with variance $2/\lambda^2$. We set $a_0=b_0=0.1$, $\tau_0^2=10$, and $\pi_0=0.05$ as suggested in \cite{zhang2021bayesian}.
Here, $\lambda$ is similar to our global variance parameter $W$, while $\gamma_j$ encodes the inclusion/exclusion of the coefficient in the model. Note that the original paper proposed this prior for the cumulative logit model, which means that we are unable to directly use the \texttt{ordinalbayes} package \citep{archer2022ordinalbayes}. Additionally, it is well-documented that the spike-and-slab prior cannot be implemented in \texttt{Stan}, so we code this model in \texttt{JAGS} \citep{plummer2016}.

Lastly, the Horseshoe prior \citep{carvalho2009handling} is a popular shrinkage prior, and can easily be adapted to the ordinal regression setting. Specifically, 
\begin{align*}
    \beta_j\mid \tau, \lambda_j &\sim \mathsf{Normal}(0,\lambda_j\tau^2)\\
    \lambda_j,\tau^2&\sim \mathsf{HalfCauchy}(1)\\   
    \tau_k&\sim\mathsf{Normal}(0,\tau_0^2)\text{ where }\tau_1<\tau_2<\cdots<\tau_{K-1}.
\end{align*}
This model is coded in \texttt{Stan} using the auxiliary variable formulation of \cite{makalic2015simple}. All experiments were conducted on a 2024 M4 Mac Mini with 16 GB of memory and running four MCMC chains in parallel.

\subsection{Results}
The results for $n=50,p=250$ and $n=100,p=500$ are in Figures \ref{tab:n50p250} and \ref{tab:n100p500}, respectively. In general, we can see that the $\mathsf{pR2D2ord}$ prior performs quite favorably across all settings. In particular, the proposed prior yields the largest AUC across all settings when the coefficients are fixed, as well as when $n=50,p=250$ and the responses are Low. Additionally, when $n=50$ and $p=250$ and the coefficients are fixed, this prior always yields the smallest MSE, and also has the lowest MSE when the coefficients are generated from the $t_3$ distribution when the response values are Low.

The original R2D2 prior also performs well, always yielding large AUC values when the coefficients are generated from the $t_3$ distribution. More generally, all three R2D2 methods have significantly larger AUC compared to ssLASSO and HS when the coefficients are from a $t_3$ distribution, and when the coefficients are fixed, they greatly outperform ssLASSO. Moreover, when the response values are Low, the R2D2 methods yield superior MSEs values to those of ssLASSO. ssLASSO shows the best MSE performance for $t_3$ and Even response variables, while HS has the best MSE for $n=100,p=500$ and $K=5$.

All methods yield coverage at or above the expected level, though $\mathsf{pR2D2ord}$ consistently has wider intervals. In Table \ref{tab:run_time}, we find that all R2D2 methods, including the proposed method, are significantly faster than both HS and ssLASSO. Indeed, when $n=100$ and $p=1000$, the proposed method is three and four times faster than HS and ssLASSO, respectively. As previously mentioned, because of the spike-and-slab framework, ssLASSO cannot be implemented in \texttt{Stan}, but instead must be run in \texttt{JAGS}. This is likely part of the reason for the difference in computation time. Recall that we introduced the auxiliary variable formulation for the GIG distribution with the express purpose of using \texttt{Stan}; this extra step resulted in a large computational advantage compared to ssLASSO and HS.

When ssLASSO or HS yields the smallest MSE or largest AUC, the proposed method's values are usually quite comporable. On the other hand, there are many situations where $\mathsf{pR2D2ord}$ has significantly better performance than these competitors. Also noteworthy is that $\mathsf{pR2D2ord}$ outperforms R2D2glm in terms of MSE and AUC in almost all settings. The R2D2glm prior is philosophically problematic since it takes the mean and variance of ordinal data. So not only is the proposed method more satisfying philosophically, it also yields superior empirical performance.

Lastly, the proposed method particularly shines when the responses are imbalanced between classes (Low) when $n=50$ and $p=250$. While we only demonstrated this for an excess of small values, we found similar trends when the data was imbalanced in different ways. As real-world responses are unlikely to be uniform across classes, this is also a major advantage of the proposed method. Moreover, if the practitioner expects that data to be skewed in a particular direction, this can be explicitly accounted for in the prior specification via $\boldsymbol{\alpha}$. In these simulations, we simply took all entries of $\boldsymbol{\alpha}$ to be the same, so utilizing this prior information should only improve the performance.

\begin{figure}
    \centering
    \begin{subfigure}[b]{\textwidth}
        \centering
    \begin{tabular}{l|cccc|cccc}
    \multicolumn{1}{l|} {} & \multicolumn{4}{|c|}{Fixed} & \multicolumn{4}{c}{$t_3$} \\  
\multicolumn{1}{l}{Prior} & \multicolumn{1}{|c}{MSE (se)} & \multicolumn{1}{c}{AUC} & \multicolumn{1}{c}{Cov.} & \multicolumn{1}{c|}{Width} & \multicolumn{1}{c}{MSE (se)} & \multicolumn{1}{c}{AUC} & \multicolumn{1}{c}{Cov.} &\multicolumn{1}{c}{Width} \\ \hline
HS & 0.24 (0.01) & 0.79 & 0.99 & 0.67 & 0.20 (0.01) & 0.72 & 0.99 & 0.58\\ 
ssLASSO & 0.24 (0.01) & 0.69 & 0.99 & 0.30 & {\bf 0.19} (0.01) & 0.63 & 0.99 & 0.23\\ 
R2D2 & 0.15 (0.00) & 0.81 & 0.98 & 0.11 & 0.21 (0.01) & {\bf 0.88} & 0.98 & 0.14\\ 
R2D2glm & 0.15 (0.00) & 0.78 & 0.98 & 0.12 & 0.21 (0.01) & 0.86 & 0.98 & 0.16\\ 
$\mathsf{pR2D2ord}$ & {\bf 0.15} (0.00) & {\bf 0.88} & 0.98 & 0.36 & 0.20 (0.01) & 0.87 & 0.98 & 0.47
    \end{tabular}
    
        \caption{$K=3$, Even}
        
    \end{subfigure}\\
    \begin{subfigure}[b]{\textwidth}
        \centering
    \begin{tabular}{l|cccc|cccc}
    \multicolumn{1}{l|} {} & \multicolumn{4}{|c|}{Fixed} & \multicolumn{4}{c}{$t_3$} \\  
\multicolumn{1}{l}{Prior} & \multicolumn{1}{|c}{MSE (se)} & \multicolumn{1}{c}{AUC} & \multicolumn{1}{c}{Cov.} & \multicolumn{1}{c|}{Width} & \multicolumn{1}{c}{MSE (se)} & \multicolumn{1}{c}{AUC} & \multicolumn{1}{c}{Cov.} &\multicolumn{1}{c}{Width} \\ \hline
HS & 0.47 (0.03) & 0.78 & 0.99 & 1.21 & 0.43 (0.03) & 0.68 & 0.99 & 1.06\\ 
ssLASSO & 0.45 (0.02) & 0.67 & 0.99 & 0.58 & 0.47 (0.04) & 0.62 & 0.99 & 0.56\\ 
R2D2 & 0.15 (0.00) & 0.80 & 0.98 & 0.10 & 0.22 (0.01) & 0.84 & 0.98 & 0.12\\ 
R2D2glm & 0.15 (0.00) & 0.77 & 0.98 & 0.12 & 0.22 (0.01) & 0.84 & 0.98 & 0.14\\ 
$\mathsf{pR2D2ord}$ & {\bf 0.15} (0.00) & {\bf 0.86} & 0.98 & 0.38 & {\bf 0.21} (0.01) & {\bf 0.86} & 0.98 & 0.43
    \end{tabular}
    
        \caption{$K=3$, Low}
        
    \end{subfigure}\\
    \begin{subfigure}[b]{\textwidth}
        \centering
    \begin{tabular}{l|cccc|cccc}
    \multicolumn{1}{l|} {} & \multicolumn{4}{|c|}{Fixed} & \multicolumn{4}{c}{$t_3$} \\  
\multicolumn{1}{l}{Prior} & \multicolumn{1}{|c}{MSE (se)} & \multicolumn{1}{c}{AUC} & \multicolumn{1}{c}{Cov.} & \multicolumn{1}{c|}{Width} & \multicolumn{1}{c}{MSE (se)} & \multicolumn{1}{c}{AUC} & \multicolumn{1}{c}{Cov.} &\multicolumn{1}{c}{Width} \\ \hline
HS & 0.17 (0.00) & 0.83 & 0.99 & 0.43 & 0.14 (0.01) & 0.72 & 0.99 & 0.34\\ 
ssLASSO & 0.16 (0.01) & 0.73 & 0.99 & 0.18 & {\bf 0.12} (0.01) & 0.66 & 0.99 & 0.14\\ 
R2D2 & 0.15 (0.00) & 0.84 & 0.98 & 0.12 & 0.20 (0.01) & {\bf 0.87} & 0.98 & 0.16\\ 
R2D2glm & 0.15 (0.00) & 0.81 & 0.98 & 0.12 & 0.20 (0.01) & 0.85 & 0.98 & 0.16\\ 
$\mathsf{pR2D2ord}$ & {\bf 0.14} (0.00) & {\bf 0.91} & 0.98 & 0.41 & 0.17 (0.01) & 0.86 & 0.98 & 0.47 
    \end{tabular}
    
        \caption{$K=5$, Even}
        
    \end{subfigure}\\

    \begin{subfigure}[b]{\textwidth}
        \centering
    \begin{tabular}{l|cccc|cccc}
    \multicolumn{1}{l|} {} & \multicolumn{4}{|c|}{Fixed} & \multicolumn{4}{c}{$t_3$} \\  
\multicolumn{1}{l}{Prior} & \multicolumn{1}{|c}{MSE (se)} & \multicolumn{1}{c}{AUC} & \multicolumn{1}{c}{Cov.} & \multicolumn{1}{c|}{Width} & \multicolumn{1}{c}{MSE (se)} & \multicolumn{1}{c}{AUC} & \multicolumn{1}{c}{Cov.} &\multicolumn{1}{c}{Width} \\ \hline
HS & 0.46 (0.02) & 0.79 & 0.99 & 1.30 & 0.47 (0.04) & 0.69 & 0.99 & 1.08\\ 
ssLASSO & 0.38 (0.02) & 0.69 & 0.99 & 0.49 & 0.36 (0.02) & 0.63 & 0.99 & 0.44\\ 
R2D2 & 0.15 (0.00) & 0.80 & 0.98 & 0.11 & 0.23 (0.01) & 0.86 & 0.98 & 0.13\\ 
R2D2glm & 0.15 (0.00) & 0.77 & 0.98 & 0.11 & 0.23 (0.01) & 0.84 & 0.98 & 0.14\\ 
$\mathsf{pR2D2ord}$ & {\bf 0.15} (0.00) & {\bf 0.87} & 0.98 & 0.35 & {\bf 0.21} (0.01) & {\bf 0.87} & 0.98 & 0.44 
    \end{tabular}
    
        \caption{$K=5$, Low}
        
    \end{subfigure}
    \caption{Simulation results for $n=50$ and $p=250$. The sub-captions specify the value of $K$ and the distribution of the responses. The first header corresponds to the distribution of $\bbeta$. The columns correspond to the mean-squared error, AUC, empirical coverage and width of the 95\% credible intervals. The lowest MSE and largest AUC for each setting are denoted in {\bf bold}.}
    \label{tab:n50p250}
\end{figure}


\begin{figure}
    \centering
    \begin{subfigure}[b]{\textwidth}
        \centering
    \begin{tabular}{l|cccc|cccc}
    \multicolumn{1}{l|} {} & \multicolumn{4}{|c|}{Fixed} & \multicolumn{4}{c}{$t_3$} \\  
\multicolumn{1}{l}{Prior} & \multicolumn{1}{|c}{MSE (se)} & \multicolumn{1}{c}{AUC} & \multicolumn{1}{c}{Cov.} & \multicolumn{1}{c|}{Width} & \multicolumn{1}{c}{MSE (se)} & \multicolumn{1}{c}{AUC} & \multicolumn{1}{c}{Cov.} &\multicolumn{1}{c}{Width} \\ \hline
HS & {\bf 0.09} (0.00) & 0.86 & 0.99 & 0.13 & {\bf 0.08} (0.01) & 0.76 & 0.99 & 0.11\\ 
ssLASSO & 0.13 (0.01) & 0.81 & 1.00 & 0.16 & 0.11 (0.01) & 0.71 & 0.99 & 0.12\\ 
R2D2 & 0.11 (0.00) & 0.91 & 0.99 & 0.09 & 0.15 (0.01) & {\bf 0.91} & 0.99 & 0.15\\ 
R2D2glm & 0.11 (0.00) & 0.90 & 0.99 & 0.11 & 0.15 (0.01) & 0.89 & 0.99 & 0.16\\ 
$\mathsf{pR2D2ord}$ & 0.10 (0.00) & {\bf 0.96} & 0.99 & 0.33 & 0.13 (0.01) & 0.90 & 0.99 & 0.40
    \end{tabular}
    
        \subcaption{$K=3$, Even}
        
    \end{subfigure}\\
    \begin{subfigure}[b]{\textwidth}
        \centering
    \begin{tabular}{l|cccc|cccc}
    \multicolumn{1}{l|} {} & \multicolumn{4}{|c|}{Fixed} & \multicolumn{4}{c}{$t_3$} \\  
\multicolumn{1}{l}{Prior} & \multicolumn{1}{|c}{MSE (se)} & \multicolumn{1}{c}{AUC} & \multicolumn{1}{c}{Cov.} & \multicolumn{1}{c|}{Width} & \multicolumn{1}{c}{MSE (se)} & \multicolumn{1}{c}{AUC} & \multicolumn{1}{c}{Cov.} &\multicolumn{1}{c}{Width} \\ \hline
HS & 0.11 (0.00) & 0.83 & 0.99 & 0.19 & {\bf 0.10} (0.00) & 0.71 & 0.99 & 0.16\\ 
ssLASSO & 0.26 (0.01) & 0.76 & 1.00 & 0.29 & 0.23 (0.01) & 0.67 & 0.99 & 0.24\\ 
R2D2 & 0.11 (0.00) & 0.89 & 0.99 & 0.09 & 0.15 (0.01) & {\bf 0.92} & 0.99 & 0.13\\ 
R2D2glm & 0.11 (0.00) & 0.87 & 0.99 & 0.10 & 0.15 (0.01) & 0.89 & 0.99 & 0.13\\ 
$\mathsf{pR2D2ord}$ & {\bf 0.10} (0.00) & {\bf 0.95} & 0.99 & 0.31 & 0.13 (0.01) & 0.90 & 0.99 & 0.39
    \end{tabular}\\
    
        \caption{$K=3$, Low}
        
    \end{subfigure}
    \begin{subfigure}[b]{\textwidth}
        \centering
    \begin{tabular}{l|cccc|cccc}
    \multicolumn{1}{l|} {} & \multicolumn{4}{|c|}{Fixed} & \multicolumn{4}{c}{$t_3$} \\  
\multicolumn{1}{l}{Prior} & \multicolumn{1}{|c}{MSE (se)} & \multicolumn{1}{c}{AUC} & \multicolumn{1}{c}{Cov.} & \multicolumn{1}{c|}{Width} & \multicolumn{1}{c}{MSE (se)} & \multicolumn{1}{c}{AUC} & \multicolumn{1}{c}{Cov.} &\multicolumn{1}{c}{Width} \\ \hline
HS & {\bf 0.07} (0.00) & 0.90 & 1.00 & 0.10 & {\bf 0.06} (0.00) & 0.77 & 0.99 & 0.08\\ 
ssLASSO & 0.08 (0.00) & 0.87 & 1.00 & 0.09 & 0.06 (0.00) & 0.74 & 1.00 & 0.07\\ 
R2D2 & 0.11 (0.00) & 0.94 & 0.99 & 0.11 & 0.12 (0.01) & {\bf 0.89} & 0.99 & 0.18\\ 
R2D2glm & 0.11 (0.00) & 0.91 & 0.99 & 0.11 & 0.12 (0.01) & 0.86 & 0.99 & 0.18\\ 
$\mathsf{pR2D2ord}$ & 0.09 (0.00) & {\bf 0.97} & 0.99 & 0.36 & 0.10 (0.00) & 0.88 & 0.99 & 0.38
    \end{tabular}\\
    
        \caption{$K=5$, Even}
        
    \end{subfigure}

    \begin{subfigure}[b]{\textwidth}
        \centering
    \begin{tabular}{l|cccc|cccc}
    \multicolumn{1}{l|} {} & \multicolumn{4}{|c|}{Fixed} & \multicolumn{4}{c}{$t_3$} \\  
\multicolumn{1}{l}{Prior} & \multicolumn{1}{|c}{MSE (se)} & \multicolumn{1}{c}{AUC} & \multicolumn{1}{c}{Cov.} & \multicolumn{1}{c|}{Width} & \multicolumn{1}{c}{MSE (se)} & \multicolumn{1}{c}{AUC} & \multicolumn{1}{c}{Cov.} &\multicolumn{1}{c}{Width} \\ \hline
HS & 0.11 (0.00) & 0.85 & 1.00 & 0.20 & {\bf 0.11} (0.01) & 0.75 & 0.99 & 0.17\\ 
ssLASSO & 0.18 (0.01) & 0.79 & 1.00 & 0.21 & 0.16 (0.01) & 0.69 & 0.99 & 0.18\\ 
R2D2 & 0.11 (0.00) & 0.87 & 0.99 & 0.09 & 0.15 (0.01) & {\bf 0.92} & 0.99 & 0.14\\ 
R2D2glm & 0.11 (0.00) & 0.86 & 0.99 & 0.09 & 0.15 (0.01) & 0.90 & 0.99 & 0.14\\ 
$\mathsf{pR2D2ord}$ & {\bf 0.10} (0.00) & {\bf 0.94} & 0.99 & 0.28 & 0.13 (0.01) & 0.89 & 0.99 & 0.38 
    \end{tabular}\\
    
        \caption{$K=5$, Low}
        
    \end{subfigure}
    \caption{Simulation results for $n=100$ and $p=500$. The sub-captions specify the value of $K$ and the distribution of the responses. The first header corresponds to the distribution of $\bbeta$. The columns correspond to the mean-squared error, AUC, empirical coverage and width of the 95\% credible intervals. The lowest MSE and largest AUC for each setting are denoted in {\bf bold}.}
    \label{tab:n100p500}
\end{figure}

\begin{table}[]
    \centering
    \begin{tabular}{c|cccc}
           & $(50,100)$ & $(50,250)$ & $(100,500)$ & $(100,1000)$\\\hline
        HS &  65 & 130 & 322 & 584 \\
        ssLASSO & 22 & 67 & 317 & 754 \\
        R2D2 & 13 & 31 & 86 & 196\\
        R2D2glm & 9 & 16 & 36 & 72 \\
        $\mathsf{pR2D2ord}$ & 11 & 29 & 80 & 188
    \end{tabular}
    \caption{Average run-time (in seconds) for each combination of $(n,p)$, averaged over all simulation study settings.}
    \label{tab:run_time}
\end{table}


\section{Real-data analysis}\label{sec:real}
In this section, we apply the proposed method to gene expression data for hepatocellular carcinoma (HCC) \citep{archer2010identifying, archer2022ordinalbayes}. The response variable is the classification of liver tissues samples from normal $(k=1)$ to hepatitis C virus (HCV) infected but no HCC $(k=2)$ to HCV infected with HCC $(k=3)$. Such a classification naturally lends itself to ordinal modeling. The covariates correspond to the gene expressions, in addition to age and cigarette usage. After pre-processing \citep{archer2022ordinalbayes}, there are $n=242$ observations and $p=2011$ covariates. 

\subsection{Inference}
We first focus on finding the most important covariates in the dataset, as well as posterior inference for $R^2_M$ and $W$. Towards this end, we fit the $\mathsf{pR2D2ord}$ model with $a=1$ and $b=10$, and use 11,000 MCMC iterations (1000 discard as burn-in) per chain with four chains. We also fit the competing methods from Section \ref{sec:sim}. Convergence diagnostics including Rhat \citep{vehtari2021rank}, effective sample size and trace plots are included in the Supplemental Materials. Note that the MCMC chains for R2D2glm and HS did not mix well, so these methods were removed from consideration.  

First, we identify the most important factors affecting the liver tissue classification. In Table \ref{tab:inf}, we report the covariates with ten largest (in absolute value) posterior mean estimates for each model. For ssLASSO, we report the ten covariates with largest posterior probability of being non-zero. Gene names were matched with ENSEMBL IDs using the Bioconductor annotation data package, and the ENSEMBLE IDs were kept if there was no match in the package. We see that the two R2D2 methods share many similar important covariates, while those chosen by ssLASSO are generally quite different. 

\begin{figure}
    \centering
    \includegraphics[width=\linewidth]{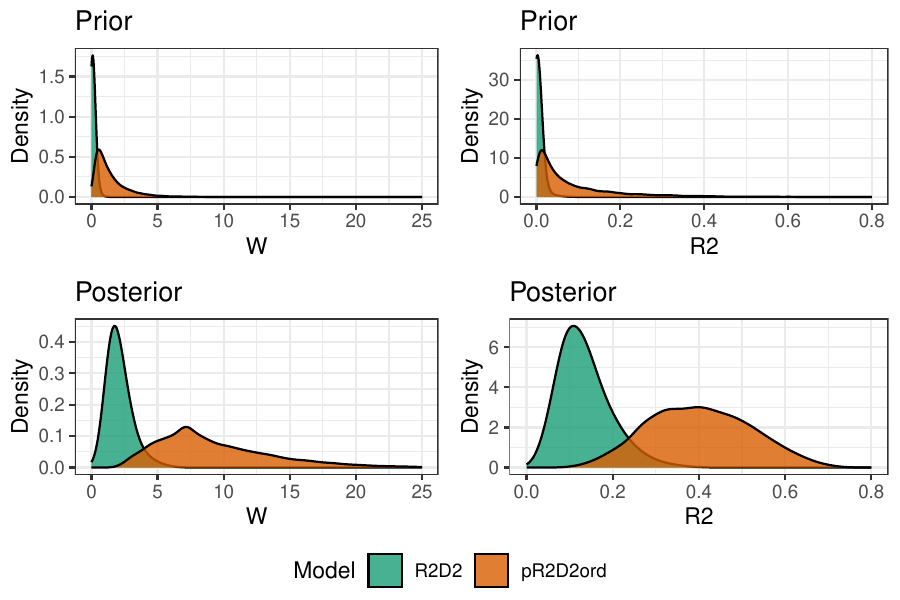}
    \caption{Prior and posterior distribution of $R^2_M$ and $W$ for the R2D2 and $\mathsf{pR2D2ord}$ priors on the gene expression data.}
    \label{fig:wR2}
\end{figure}

We also plot the prior and posterior distributions of $R^2_M$ and $W$ for each of the R2D2 methods in Figure \ref{fig:wR2}. For the proposed method, the prior distribution of $W$ has much larger mass at high values of $W$, which induces a larger prior distribution on $R^2_M$, and larger posterior distributions of $R^2_M$ and $W$, when compared with the R2D2 prior. Note that even though both R2D2 priors use $(a,b)=(1,10)$, these correspond to different definitions of $R^2$. In particular, for the R2D2 prior, this induces a beta distribution on the $R^2$ of the latent mean $\eta$, so when we compute the prior distribution of McFadden's $R^2_M$, it will not necessarily have the same distribution. Indeed, the prior distribution of $R^2_M$ for the R2D2 prior has too much mass near 0 compared to the expected $\mathsf{Beta}(1,10)$ distribution.

\begin{table}[]
    \centering
    \begin{tabular}{l|ccc}
        \multicolumn{1}{l|} {} & \multicolumn{3}{|c}{Prior}\\  
        Gene & ssLASSO & R2D2 & $\mathsf{pR2D2ord}$ \\\hline
        CWC22 & $\circ$ & &\\ 
        LINC00924 & $\circ$ & &\\ 
        KRT85 & $\circ$ & &\\ 
        HS3ST3B1 & $\circ$ & &\\ 
        ENSG233996 & $\circ$ & &\\ 
        LINC01563 & $\circ$ & &\\ 
        CCDC102B & $\circ$ & &\\ 
        ENSG177173 & $\circ$ & &\\ 
        ENSG253923 & $\circ$ & $\circ$ & $\circ$\\ 
        ENSG272071 & $\circ$ & &\\  
        ENSG230201 & & $\circ$ & $\circ$ \\ 
        ENSG203601 & & $\circ$ & $\circ$ \\ 
        ENSG260484 & & $\circ$ & $\circ$ \\ 
        MTCO1P40 & & $\circ$ & $\circ$\\ 
        CAPN6 & & $\circ$ & $\circ$\\ 
        ENSG265579 &$\circ$ \\ 
        PNN-AS1 & & $\circ$ & $\circ$ \\ 
        ENSG250602 & & $\circ$ & $\circ$ \\ 
        ENSG271711 & & $\circ$ & $\circ$\\ 
        ENSG263612 &  & & $\circ$
    \end{tabular}
    \caption{Inference results on gene expression data. A $\circ$ indicates that this covariate was one of the ten largest (in absolute value) posterior mean estimates for each model. For ssLASSO, it corresponds to the largest posterior probability of being non-zero.}
    \label{tab:inf}
\end{table}

\subsection{Prediction}
We also study the predictive performance of the proposed method. Specifically, we split the dataset via an 80/20 training/testing split, and fit the model on the training data to predict the response on the out-of-sample testing data. Given the posterior means $\hat\bbeta$ and $\hat{\boldsymbol{\tau}}$, we compute $\hat{\boldsymbol{\eta}}={\bf X}_{\text{test}}\hat{\bbeta}$ where ${\bf X}_{\text{test}}$ are the covariates for the testing data. Then for testing data $i$, we find
$$
    \hat Y_i
    =\begin{cases}
        1&\hat\eta_i<\hat\tau_1\\
        2&\hat\tau_1\leq \hat\eta_i\leq \hat\tau_2\\
        3& \hat\tau_3<\hat\eta_i
    \end{cases}
    .
$$
Given the predictions $\hat{\bf Y}$ and testing data ${\bf Y}_{\text{test}}$, we compute the accuracy and root mean squared error (MSE) of the predictions as
$$
    \text{acc}
    =\frac{1}{\tilde n}\sum_{i=1}^{\tilde n}\mathbb I(\hat Y_i=Y_{i,\text{test}})
\ \ \ \ \text{ and }\ \ \ \ 
    \text{mse}
    =\sqrt{\frac{1}{\tilde n}\sum_{i=1}^{\tilde n}(\hat Y_i-Y_{i,\text{test}})^2},
$$
where $\mathbb I(\cdot)$ is the indicator function and $\tilde n$ is the number of observations in the testing data. Note that MSE is not strictly appropriate since the response is ordinal and this measure relies on the mean, but we are employing this metric as a general measure of model fit. For each method, we use 4000 MCMC iterations per chain (1000 burn-in, four chains), compute the accuracy and MSE, and take the average over 50 MC replicates. The results are in Table \ref{tab:pred}. The $\mathsf{pR2D2ord}$ prior yields the highest accuracy, as well as the lowest MSE, demonstrating good predictive performance.

\begin{table}[]
    \centering
    \begin{tabular}{c|cc}
        Prior & Acc. (se) & MSE (se) \\\hline
        ssLASSO & 0.61 (0.01) & 0.71 (0.01)\\
        R2D2 & 0.62 (0.01) & 0.68 (0.01)\\
        $\mathsf{pR2D2ord}$ & {\bf 0.65} (0.01)& {\bf 0.65} (0.01) 
    \end{tabular}
    \caption{Accuracy (Acc.) and mean squared error (MSE) prediction results for real-data analysis with standard error in parentheses. The highest accuracy and lowest MSE are in {\bf bold}.}
    \label{tab:pred}
\end{table}

\section{Conclusion}\label{sec:conc}

In this work, we propose the $\mathsf{pR2D2ord}$ prior for high-dimensional ordinal regression. We adopted the R2D2 paradigm, but extended it to a different definition of $R^2$. By assigning a generalized inverse Gaussian prior to the global variance parameter, we showed that this induces a beta prior distribution on $R^2_M$, and an auxiliary variable trick allowed this prior to be implemented in \texttt{Stan}. Not only is the proposed method philosophically satisfying, it also yields excellent empirical performance on both simulated and real-world data. Moreover, it provides two different ways to intuitively incorporate domain knowledge. The hyperparameters for the cut-points prior distribution, and the $a,b$ in $R^2_M\sim\mathsf{Beta}(a,b)$ allow practitioners to incorporate domain knowledge on the expected response distribution and overall model fit. In the absence of such prior information, the $\mathsf{pR2D2ord}$ prior also allows for automatic approaches for selecting hyperparameters by shrinking the model fit towards a base model.

The R2D2 prior \citep{zhang2021bayesian} has already proven to be a very promising Bayesian framework. In this work, we extend this paradigm to a new measure of model fit, McFadden's $R^2$. This demonstrates that the R2D2 paradigm can be extended to even more situations, opening up new avenues for creative ways to set priors on model fit.

\section*{Funding}
This work was supported by JSPS KAKENHI grant
24K22613.

\bibliographystyle{apalike}
\bibliography{refs}

\clearpage

\begin{center}
    \Large Supplemental Materials
\end{center}

\section*{Derivation of distribution of $\boldsymbol{\tau}$}
{\it Proof.} Assume that $\boldsymbol{\pi}=(\pi_1,\dots,\pi_K)\sim\mathsf{Dirichlet}(\alpha_1,\dots,\alpha_K)$. Note that $\pi_K=1-\sum_{k=1}^{K-1}\pi_k$ so we can exclude this from the transformation. Then we want to find the distribution of the cut-points, $\boldsymbol{\tau}=(\tau_1,\dots,\tau_{K-1})^T$ where $\boldsymbol{\pi}$ and $\boldsymbol{\tau}$ are related by
$$
\pi_k(\boldsymbol{\tau})
=
\begin{cases}
\Phi_W(\tau_1)&k=1\\
\Phi_W(\tau_k)-\Phi_W(\tau_{k-1})&k=2,\dots,K-1
\end{cases}
$$
and $\Phi_W(\cdot)$ is the cumulative distribution function for a normal random variable with mean 0 and variance $1+W$. We can easily show that
$$
    \frac{\partial \pi_k(\boldsymbol{\tau})}{\partial \tau_j}
    =
    \begin{cases}
        \phi_W(\tau_j)&k=j\\
        -\phi_W(\tau_j)&k=j-1\\
        0&\text{otherwise}
    \end{cases}
$$
where $\phi_W(\cdot)$ is the probability distribution function for a normal random variable with mean 0 and variance $1+W$. We can use this to construct the Jacobian matrix $J$ where
$$
    J_{jk}
    =\frac{\partial \pi_k(\boldsymbol{\tau})}{\partial \tau_j}.
$$
Thus, the induced distribution for $\boldsymbol{\tau}$ is
$$
    f(\boldsymbol{\tau})
    =d(\boldsymbol{\pi}(\boldsymbol{\tau});\boldsymbol{\alpha})\times|J|,\ \tau_1<\tau_2<\dots<\tau_{K-1}.
$$
where $d(\cdot;\boldsymbol{\alpha})$ is the pdf of the Dirichlet distribution with concentration parameters $\boldsymbol{\alpha}=(\alpha_1,\dots,\alpha_K)^T$.

\clearpage

\section*{Optimization hyperparameter values}

In Table \ref{tab:params}, we report the optimal hyperparameter values such that if $W\sim\mathsf{GIG}(\lambda^*,\rho^*,\chi^*$), then $R^2_M\stackrel{\text{approx.}}{\sim}\mathsf{Beta}(a,b)$ for various values of $n$ and $K$. For each combination of parameter values, we set $\boldsymbol{\alpha}=(1,1,\dots,1)^T$, run the optimization routine five times, and report the parameters yielding the lowest value of the objective function.

\begin{table}[h]
    \centering
    \begin{tabular}{c|ccccccccccccc}
    \multicolumn{1}{c|} {$(a,b)$} &\multicolumn{4}{c} {$(1,1)$} & \multicolumn{4}{c}{$(1,5)$} & \multicolumn{4}{c}{$(1,10)$} \\ 
    \multicolumn{1}{c|} {$n$} &\multicolumn{2}{c} {$100$} & \multicolumn{2}{c}{$1000$} &\multicolumn{2}{c} {$100$} &  \multicolumn{2}{c}{$1000$}&\multicolumn{2}{c} {$100$}   & \multicolumn{2}{c}{$1000$}\\ 
    \multicolumn{1}{c|} {$K$} &\multicolumn{1}{c} {$3$} &  \multicolumn{1}{c}{$5$} &\multicolumn{1}{c} {$3$} & \multicolumn{1}{c}{$5$} &\multicolumn{1}{c} {$3$} &  \multicolumn{1}{c}{$5$} &\multicolumn{1}{c} {$3$} & \multicolumn{1}{c}{$5$} &\multicolumn{1}{c} {$3$} &  \multicolumn{1}{c}{$5$} 
    &\multicolumn{1}{c} {$3$} &  \multicolumn{1}{c}{$5$}\\ 
    \hline
    $\lambda^*$ & 1.10 &0.67 &1.23 &0.85 &0.67 &0.51 &0.36 &0.38 &0.01 &0.05 &0.00 &0.06\\
    $\rho^*$ & 1.41 &3.76 &2.64 &1.80 &1.19 &1.56 &1.5 &1.39 &1.00 &1.08 &1.00 &0.97\\
    $\chi^*$ & 0.15 &0.14 &0.19 &0.14 &0.77 &0.70 &0.65 &0.66 &1.04 &1.03 &1.00 &0.99 
    \end{tabular}
    \caption{Optimal values of $\lambda^*,\rho^*,\chi^*$ for various values of $a,b,n$ and $K$.}
    \label{tab:params}
\end{table}

\section*{Beta distribution approximation}
\begin{figure}
    \centering
    \includegraphics[width=\linewidth]{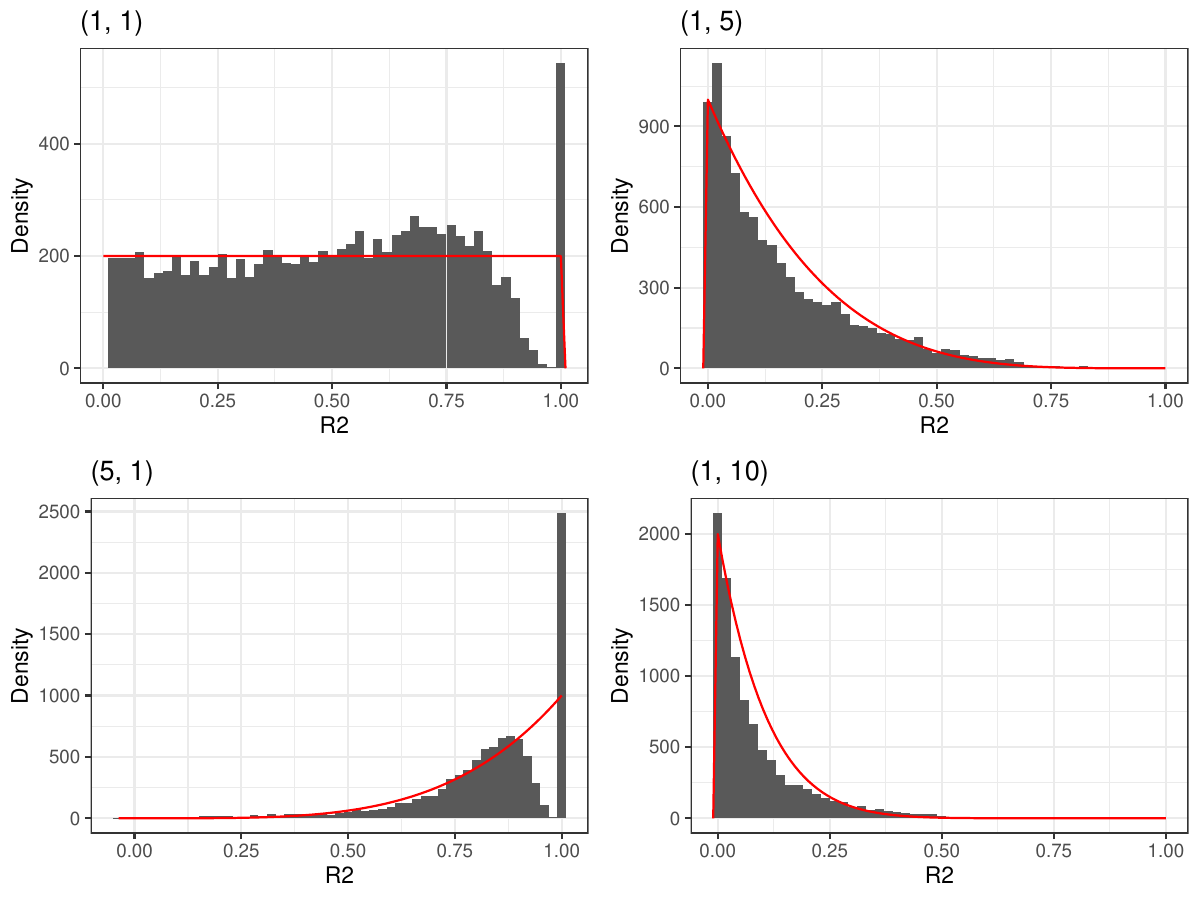}
    \caption{Empirical distribution of $R^2_M$ compared against the theoretical distribution. The panel's title corresponds to the value of $(a,b)$.}
    \label{fig:hist}
\end{figure}

In this section, we show the closeness of the approximation of the empirical distribution of $R^2_M$ (when using the optimal hyperparameters), with the desired beta distribution. In particular, we set $n=100$, $K=5$ and $\boldsymbol{\alpha}=(1,1,\dots, 1)^\top$, and vary the values of $a$ and $b$. After using the optimization procedure to select the hyperparameters for $W$, we generate values of $R^2_M$ given the following procedure:
\begin{enumerate}
    \item Sample $W\sim\mathsf{GIG}(\lambda^*,\rho^*,\chi^*)$
    \item Sample $\boldsymbol{\tau}\sim f(\boldsymbol{\tau};\boldsymbol{\alpha})$ from (8)
    \item Sample $\tilde Y_1,\dots,\tilde Y_n\mid W\stackrel{\text{iid.}}{\sim}\mathsf{Normal}(0,1+W)$
    \item Compute ${\bf Y}$ using (3)
    \item Compute $R^2_M$ using (11)
\end{enumerate}
This process is repeated $N=10,000$ times to obtain a distribution of $R^2_M$. We plot the histogram of this distribution against the true beta distribution pdf (in red) to assess the approximation. The plots for $(a,b)\in\{(1,1), (1,5), (1,10), (5,1)\}$ are in Figure \ref{fig:hist}. We can see that the fit is quite good for priors with large mass near 0, i.e., $(1,5)$ and $(1,10)$. It is more difficult to generate large values of $R^2_M$ close to but smaller than 1, however, as evidenced by the examples with $(1,1)$ and $(5,1)$. In both of these cases, the empirical distribution is quite close to the theoretical distribution, except for at $R^2_M=1$. In general, this is not concerning as we prefer prior distributions for $R^2_M$ with large prior mass near 0, as this enforces sparsity in the model. Additionally, the R2D2 paradigm is simply being used to select hyperparameters for the prior distribution, so a perfect fit is not required.

\clearpage

\section*{Simulation Results}

In Figures \ref{tab:n50p100} and \ref{tab:n100p1000}, we report additional results from the simulation study. The details of these settings can be found in the main manuscript. The results are similar to those reported in the main manuscript.

\begin{figure}
    \centering
    \begin{subfigure}[b]{\textwidth}
        \centering
    \begin{tabular}{l|cccc|cccc}
    \multicolumn{1}{l|} {} & \multicolumn{4}{|c|}{Fixed} & \multicolumn{4}{c}{$t_3$} \\  
\multicolumn{1}{l}{Prior} & \multicolumn{1}{|c}{MSE (se)} & \multicolumn{1}{c}{AUC} & \multicolumn{1}{c}{Cov.} & \multicolumn{1}{c|}{Width} & \multicolumn{1}{c}{MSE (se)} & \multicolumn{1}{c}{AUC} & \multicolumn{1}{c}{Cov.} &\multicolumn{1}{c}{Width} \\ \hline
HS & 0.60 (0.03) & 0.85 & 0.99 & 2.13 & 0.45 (0.03) & 0.76 & 0.99 & 1.55\\ 
ssLASSO & 0.23 (0.01) & 0.70 & 0.98 & 0.23 & {\bf 0.20} (0.01) & 0.64 & 0.97 & 0.21\\ 
R2D2 & 0.24 (0.00) & 0.86 & 0.94 & 0.21 & 0.33 (0.02) & {\bf 0.87} & 0.95 & 0.28\\ 
R2D2glm & 0.23 (0.00) & 0.86 & 0.94 & 0.23 & 0.32 (0.02) & 0.86 & 0.95 & 0.32\\ 
pR2D2ord & {\bf 0.20} (0.00) & {\bf 0.93} & 0.96 & 0.66 & 0.27 (0.02) & 0.83 & 0.97 & 0.72
    \end{tabular}
    
        \caption{$K=3$, Even}
        
    \end{subfigure}\\
    \begin{subfigure}[b]{\textwidth}
        \centering
    \begin{tabular}{l|cccc|cccc}
    \multicolumn{1}{l|} {} & \multicolumn{4}{|c|}{Fixed} & \multicolumn{4}{c}{$t_3$} \\  
\multicolumn{1}{l}{Prior} & \multicolumn{1}{|c}{MSE (se)} & \multicolumn{1}{c}{AUC} & \multicolumn{1}{c}{Cov.} & \multicolumn{1}{c|}{Width} & \multicolumn{1}{c}{MSE (se)} & \multicolumn{1}{c}{AUC} & \multicolumn{1}{c}{Cov.} &\multicolumn{1}{c}{Width} \\ \hline
HS & 2.70 (0.22) & 0.82 & 0.99 & 8.26 & 4.11 (0.82) & 0.75 & 0.99 & 8.46\\ 
ssLASSO & 0.47 (0.04) & 0.68 & 0.98 & 0.56 & 0.65 (0.09) & 0.62 & 0.97 & 0.68\\ 
R2D2 & 0.24 (0.00) & 0.84 & 0.94 & 0.19 & 0.34 (0.02) & {\bf 0.85} & 0.95 & 0.24\\ 
R2D2glm & 0.24 (0.00) & 0.83 & 0.94 & 0.23 & 0.34 (0.02) & 0.84 & 0.95 & 0.27\\ 
pR2D2ord & {\bf 0.21} (0.00) & {\bf 0.90} & 0.96 & 0.67 & {\bf 0.30} (0.02) & 0.82 & 0.96 & 0.69
    \end{tabular}\\
    
        \caption{$K=3$, Low}
        
    \end{subfigure}
    \begin{subfigure}[b]{\textwidth}
        \centering
    \begin{tabular}{l|cccc|cccc}
    \multicolumn{1}{l|} {} & \multicolumn{4}{|c|}{Fixed} & \multicolumn{4}{c}{$t_3$} \\  
\multicolumn{1}{l}{Prior} & \multicolumn{1}{|c}{MSE (se)} & \multicolumn{1}{c}{AUC} & \multicolumn{1}{c}{Cov.} & \multicolumn{1}{c|}{Width} & \multicolumn{1}{c}{MSE (se)} & \multicolumn{1}{c}{AUC} & \multicolumn{1}{c}{Cov.} &\multicolumn{1}{c}{Width} \\ \hline
HS & 0.29 (0.01) & 0.89 & 0.98 & 0.96 & 0.21 (0.01) & 0.78 & 0.98 & 0.72\\ 
ssLASSO & 0.19 (0.00) & 0.75 & 0.98 & 0.17 & {\bf 0.17} (0.01) & 0.68 & 0.97 & 0.13\\ 
R2D2 & 0.23 (0.00) & 0.91 & 0.94 & 0.24 & 0.29 (0.01) & {\bf 0.86} & 0.95 & 0.32\\ 
R2D2glm & 0.23 (0.00) & 0.90 & 0.94 & 0.26 & 0.28 (0.01) & 0.85 & 0.95 & 0.34\\ 
pR2D2ord & {\bf 0.18} (0.00) & {\bf 0.96} & 0.98 & 0.73 & 0.23 (0.01) & 0.85 & 0.97 & 0.74 
    \end{tabular}\\
    
        \caption{$K=5$, Even}
        
    \end{subfigure}

    \begin{subfigure}[b]{\textwidth}
        \centering
    \begin{tabular}{l|cccc|cccc}
    \multicolumn{1}{l|} {} & \multicolumn{4}{|c|}{Fixed} & \multicolumn{4}{c}{$t_3$} \\  
\multicolumn{1}{l}{Prior} & \multicolumn{1}{|c}{MSE (se)} & \multicolumn{1}{c}{AUC} & \multicolumn{1}{c}{Cov.} & \multicolumn{1}{c|}{Width} & \multicolumn{1}{c}{MSE (se)} & \multicolumn{1}{c}{AUC} & \multicolumn{1}{c}{Cov.} &\multicolumn{1}{c}{Width} \\ \hline
HS & 1.70 (0.11) & 0.83 & 0.99 & 5.60 & 2.70 (0.37) & 0.74 & 0.99 & 6.00\\ 
ssLASSO & 0.40 (0.02) & 0.70 & 0.98 & 0.45 & 0.46 (0.04) & 0.64 & 0.97 & 0.47\\ 
R2D2 & 0.24 (0.00) & 0.85 & 0.94 & 0.21 & 0.35 (0.02) & {\bf 0.87} & 0.95 & 0.27\\ 
R2D2glm & 0.24 (0.00) & 0.84 & 0.94 & 0.22 & 0.35 (0.02) & 0.86 & 0.95 & 0.29\\ 
pR2D2ord & {\bf 0.20} (0.00) & {\bf 0.93} & 0.96 & 0.64 & {\bf 0.31} (0.02) & 0.85 & 0.97 & 0.70 
    \end{tabular}\\
    
        \caption{$K=5$, Low}
        
    \end{subfigure}
    \caption{Simulation results for $n=50$ and $p=100$. The sub-captions specify the value of $K$ and the distribution of the responses. The first header corresponds to the distribution of $\bbeta$. The columns correspond to the mean-squared error, AUC, empirical coverage and width of the 95\% credible intervals. The lowest MSE and largest AUC for each setting are denoted in {\bf bold}.}
    \label{tab:n50p100}
\end{figure}


\begin{figure}
    \centering
    \begin{subfigure}[b]{\textwidth}
        \centering
    \begin{tabular}{l|cccc|cccc}
    \multicolumn{1}{l|} {} & \multicolumn{4}{|c|}{Fixed} & \multicolumn{4}{c}{$t_3$} \\  
\multicolumn{1}{l}{Prior} & \multicolumn{1}{|c}{MSE (se)} & \multicolumn{1}{c}{AUC} & \multicolumn{1}{c}{Cov.} & \multicolumn{1}{c|}{Width} & \multicolumn{1}{c}{MSE (se)} & \multicolumn{1}{c}{AUC} & \multicolumn{1}{c}{Cov.} &\multicolumn{1}{c}{Width} \\ \hline
HS & {\bf 0.05} (0.00)  & 0.84 & 1.00 & 0.05 & {\bf 0.06} (0.00) & 0.73 & 1.00 & 0.05\\ 
ssLASSO & 0.13 (0.00)  & 0.77 & 1.00 & 0.17 & 0.10 (0.00) & 0.67 & 1.00 & 0.13\\ 
R2D2 & 0.08 (0.00)  & 0.86 & 0.99 & 0.06 & 0.10 (0.00)  & {\bf 0.93} & 0.99 & 0.09\\ 
R2D2glm & 0.08 (0.00)  & 0.83 & 0.99 & 0.07 & 0.10 (0.00)  & 0.89 & 0.99 & 0.09\\ 
pR2D2ord & 0.07 (0.00)  & {\bf 0.93} & 0.99 & 0.19 & 0.09 (0.00)  & 0.92 & 1.00 & 0.29
    \end{tabular}
    
        \caption{$K=3$, Even}
        
    \end{subfigure}\\
    \begin{subfigure}[b]{\textwidth}
        \centering
    \begin{tabular}{l|cccc|cccc}
    \multicolumn{1}{l|} {} & \multicolumn{4}{|c|}{Fixed} & \multicolumn{4}{c}{$t_3$} \\  
\multicolumn{1}{l}{Prior} & \multicolumn{1}{|c}{MSE (se)} & \multicolumn{1}{c}{AUC} & \multicolumn{1}{c}{Cov.} & \multicolumn{1}{c|}{Width} & \multicolumn{1}{c}{MSE (se)} & \multicolumn{1}{c}{AUC} & \multicolumn{1}{c}{Cov.} &\multicolumn{1}{c}{Width} \\ \hline
HS & {\bf 0.06} (0.00) & 0.81 & 1.00 & 0.06 & {\bf 0.08} (0.01) & 0.73 & 1.00 & 0.06\\ 
ssLASSO & 0.18 (0.01) & 0.73 & 1.00 & 0.26 & 0.17 (0.01) & 0.65 & 1.00 & 0.22\\ 
R2D2 & 0.08 (0.00) & 0.86 & 0.99 & 0.05 & 0.11 (0.01) & 0.92 & 0.99 & 0.08\\ 
R2D2glm & 0.08 (0.00) & 0.83 & 0.99 & 0.07 & 0.11 (0.01) & 0.90 & 0.99 & 0.08\\ 
pR2D2ord & 0.08 (0.00) & {\bf 0.93} & 0.99 & 0.18 & 0.10 (0.01) & {\bf 0.92} & 1.00 & 0.27 

    \end{tabular}\\
    
        \caption{$K=3$, Low}
        
    \end{subfigure}
    \begin{subfigure}[b]{\textwidth}
        \centering
    \begin{tabular}{l|cccc|cccc}
    \multicolumn{1}{l|} {} & \multicolumn{4}{|c|}{Fixed} & \multicolumn{4}{c}{$t_3$} \\  
\multicolumn{1}{l}{Prior} & \multicolumn{1}{|c}{MSE (se)} & \multicolumn{1}{c}{AUC} & \multicolumn{1}{c}{Cov.} & \multicolumn{1}{c|}{Width} & \multicolumn{1}{c}{MSE (se)} & \multicolumn{1}{c}{AUC} & \multicolumn{1}{c}{Cov.} &\multicolumn{1}{c}{Width} \\ \hline
 HS & {\bf 0.05} (0.00) & 0.88 & 1.00 & 0.05 & {\bf 0.05} (0.00) & 0.74 & 1.00 & 0.04\\ 
ssLASSO & 0.08 (0.00) & 0.83 & 1.00 & 0.11 & 0.07 (0.00) & 0.70 & 1.00 & 0.09\\ 
R2D2 & 0.08 (0.00) & 0.88 & 0.99 & 0.06 & 0.11 (0.01) & 0.88 & 0.99 & 0.10\\ 
R2D2glm & 0.08 (0.00) & 0.87 & 0.99 & 0.07 & 0.11 (0.01) & 0.87 & 0.99 & 0.09\\ 
pR2D2ord & 0.07 (0.00) & {\bf 0.95} & 0.99 & 0.19 & 0.10 (0.01) & {\bf 0.91} & 1.00 & 0.27   
    \end{tabular}\\

        \caption{$K=5$, Even}
        
    \end{subfigure}

    \begin{subfigure}[b]{\textwidth}
        \centering
    \begin{tabular}{l|cccc|cccc}
    \multicolumn{1}{l|} {} & \multicolumn{4}{|c|}{Fixed} & \multicolumn{4}{c}{$t_3$} \\  
\multicolumn{1}{l}{Prior} & \multicolumn{1}{|c}{MSE (se)} & \multicolumn{1}{c}{AUC} & \multicolumn{1}{c}{Cov.} & \multicolumn{1}{c|}{Width} & \multicolumn{1}{c}{MSE (se)} & \multicolumn{1}{c}{AUC} & \multicolumn{1}{c}{Cov.} &\multicolumn{1}{c}{Width} \\ \hline
HS & {\bf 0.06} (0.00) & 0.83 & 1.00 & 0.06 & {\bf 0.07} (0.00) & 0.73 & 1.00 & 0.06\\ 
ssLASSO & 0.17 (0.01) & 0.76 & 1.00 & 0.23 & 0.12 (0.00) & 0.66 & 1.00 & 0.19\\ 
R2D2 & 0.08 (0.00) & 0.89 & 0.99 & 0.05 & 0.12 (0.01) & 0.88 & 0.99 & 0.08\\ 
R2D2glm & 0.08 (0.00) & 0.81 & 0.99 & 0.06 & 0.12 (0.01) & 0.87 & 0.99 & 0.08\\ 
pR2D2ord & 0.08 (0.00) & {\bf 0.92} & 0.99 & 0.16 & 0.11 (0.01) & {\bf 0.90} & 1.00 & 0.26 
    \end{tabular}\\
        \caption{$K=5$, Low}    
    \end{subfigure}
    \caption{Simulation results for $n=100$ and $p=1000$. The sub-captions specify the value of $K$ and the distribution of the responses. The first header corresponds to the distribution of $\bbeta$. The columns correspond to the mean-squared error, AUC, empirical coverage and width of the 95\% credible intervals. The lowest MSE and largest AUC for each setting are denoted in {\bf bold}.}
    \label{tab:n100p1000}
\end{figure}

\section*{MCMC convergence in real-data analysis}

\begin{figure}
    \centering
    \includegraphics[width=\linewidth]{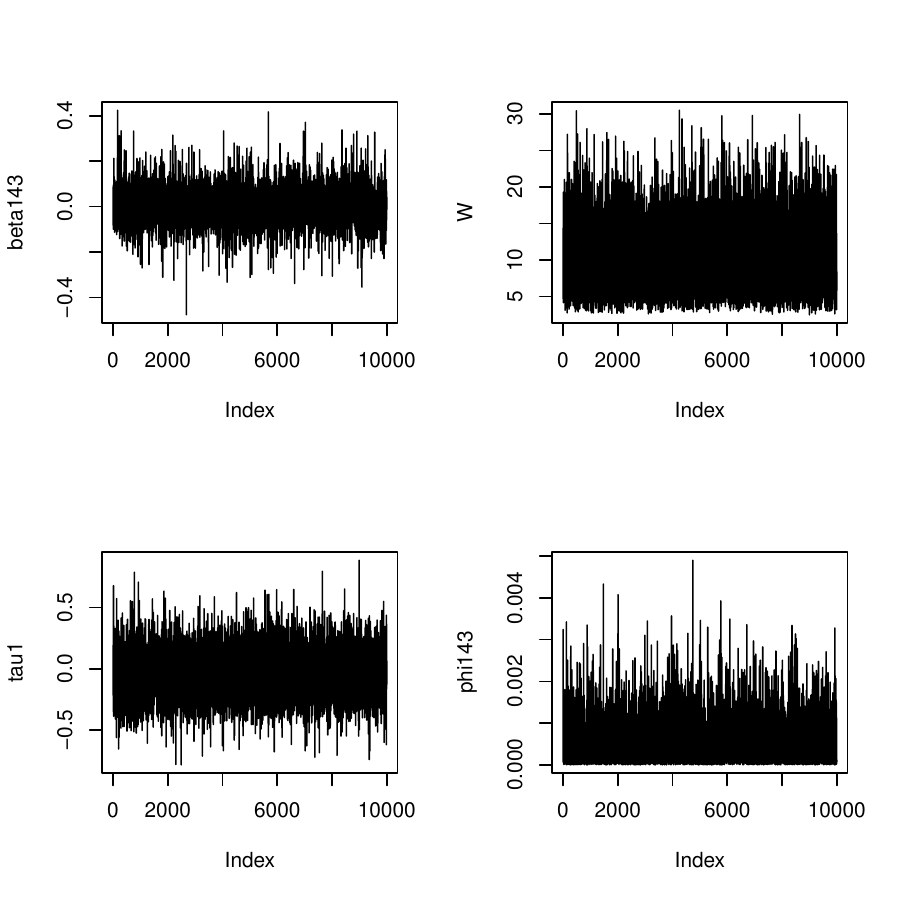}
    \caption{Trace plots for a single chain of the $\mathsf{pR2D2ord}$ model.}
    \label{fig:trace}
\end{figure}

For the $\mathsf{pR2D2ord}$ prior, we report the Rhat and effective sample size (ESS) for each of the main model parameters in Table \ref{tab:ess}. Rhat is a measure of convergence of the MCMC chains where values less than 1.05 are needed to confidently perform inference \citep{vehtari2021rank}. All values are close to 1 which indicates good mixing. We also averaged the ESS values across chains and parameters in that group. Since each chain had 10,000 MCMC samples, and the ESS values are close to 10,000, this also gives us evidence of good mixing. Finally, in Figure \ref{fig:trace}, we report trace plots for some of the main parameters in the model for a single chain. Again, we see evidence of convergence.

\begin{table}[h]
    \centering
    \begin{tabular}{c|cc}
        Parameter & Rhat &  ESS \\\hline
        $\bbeta$ & 1.00 & 9854 \\
        $W$ & 1.01 & 10087\\
        $\boldsymbol{\tau}$ & 1.01 & 9882 \\
        $\boldsymbol{\phi}$ & 1.00 & 9846
    \end{tabular}
    \caption{Rhat and effective sample size across four chains for the $\mathsf{pR2D2ord}$ prior. Each chain contains 10,000 MCMC samples.}
    \label{tab:ess}
\end{table}

\end{document}